\documentclass[twocolumn,aps,prc,superscriptaddress,showpacs,floatfix]{revtex4}

\usepackage{graphicx}
\usepackage{amssymb,amsmath}

\begin{document}

\title{Probing Nuclear Matter with Jet Conversions}

\author{W. Liu}
\affiliation{Cyclotron Institute and Department of Physics, Texas A$\&$M
University, \\ College Station, TX 77843}
\author{R. J Fries}
\affiliation{Cyclotron Institute and Department of Physics, Texas A$\&$M
University, \\ College Station, TX 77843}
\affiliation{RIKEN/BNL Research Center, Brookhaven National Laboratory,
Upton, NY 11973}

\begin{abstract}
We discuss the flavor of leading jet partons as a valuable probe of
nuclear matter. We point out that the coupling of jets to nuclear matter
naturally leads to an alteration of jet chemistry
even at high transverse momentum $p_T$. In particular,
QCD jets coupling to a chemically
equilibrated quark gluon plasma in nuclear collisions, will lead to
hadron ratios at high transverse momentum $p_T$ that can differ significantly
from their counterparts in $p+p$ collisions. Flavor measurements could
complement energy loss as a way to study interactions of hard QCD jets with
nuclear matter. Roughly speaking they probe the inverse mean free path
$1/\lambda$, while energy loss probes the average squared momentum transfer
$\mu^2/\lambda$. We present some estimates for the rate of jet conversions
in a consistent Fokker-Planck framework and their impact on future
high-$p_T$ identified hadron measurements at RHIC and LHC. We also
suggest some novel observables to test flavor effects.
\end{abstract}


\maketitle

\section{Introduction}

A hot and dense state of matter, called quark gluon plasma (QGP),
is believed to have formed during the first few microseconds after the big bang. Its degrees of freedom above a critical temperature $T_c$ consist of deconfined quarks and gluons. The Relativistic Heavy Ion Collider (RHIC) was built to study the formation and properties of quark gluon plasma in collisions of heavy ions at very large energies. We have overwhelming evidence from the first few years of running of RHIC that the temperatures reached are well above $T_c$ and that the new matter is indeed partonic \cite{whitepaper:05}.

One of the key observations at RHIC is the suppression of hadrons with large
transverse momentum $p_T$ \cite{adcox,adler1}, most impressively manifested
in reduced nuclear modification factors $R_{AA} \approx 0.2$ for hadrons in
the most central collisions for $p_T > 5$ GeV/$c$. $R_{AA}$ measures the ratio
of yields in nuclear collisions ($A+A$) vs nucleon-nucleon collisions ($N+N$)
at the same
energy scaled with the appropriate number of binary collisions. High-$p_T$
hadrons above 5 GeV/$c$ in $A+A$ collisions are believed to come
predominantly from fragmentation off QCD jets. QCD jets originate from
high-$p_T$ partons created in the initial hard collisions between the
nucleons in the nuclei. The jet quenching phenomenon at
RHIC has been attributed to loss of energy of the high-$p_T$ partons on
their way through the medium
\cite{BDMPS:96,Zakharov:96,gyulassy,wiedemann,wang,AMY:02}.
It is often characterized by a transport coefficient $\hat q = \mu^2/\lambda$,
a ratio of the average squared momentum transfer $\mu^2$ per collision and
the mean free path $\lambda$.

Initially, energy loss had been attributed to induced gluon radiation
by scattering of the leading jet parton with thermal partons from the
medium. Perturbative calculations of this process yield results
for $R_{AA}$ compatible with data after fixing one parameter characterizing
the strength of the interaction, the transport coefficient $\hat q$ or
an equivalent quantity. However, recently several inconsistencies have led
to considerable efforts to revisit this topic. First, the values for
the transport coefficients extracted from measurements of $R_{AA}$ have
large uncertainties and differ between the various theoretical models
\cite{Samu:07}. Second, the description of observables beyond those based
on the simple single inclusive spectrum is far more problematic.
As an example we mention the azimuthal anisotropy $v_2$ at high $p_T$
which is larger than predicted by theory \cite{Shuryak:01}. Third,
the application of radiative, perturbative energy loss to heavy quarks
leads to the prediction of much smaller quenching than for light quarks,
in contradiction with data from RHIC
\cite{STARcharm:06}.

The partial failure of radiative, perturbative energy loss led to renewed
interest in energy loss from elastic collisions \cite{mustafa, djordjevic1,
weiliu}, although a self-consistent treatment of elastic and radiative
energy loss in one proper unified theoretical framework is still under
construction. It has also ushered in a period of renewed interest in
non-perturbative mechanisms. Examples are energy loss through bound states in
the plasma \cite{vHR:04} --- successfully applied to heavy quarks --- and energy loss in the infinite coupling limit of QCD, modeled through the AdS/CFT correspondence \cite{adscft:06}.

In this work, we discuss the potential of flavor dependent measurements of jet
quenching observables. It is obvious that the interaction of a jet with the
medium can change the flavor of the jet --- defined here as the flavor of the
leading parton. We will show that this leads to observable differences
in the final jet hadron chemistry. Our generalized definition of flavor here includes light
quarks $q$ ($u$,$d$,$s$), gluons $g$, and photons $\gamma$.
Heavy flavors $Q$ ($c$ and $b$) should also be included.
There is no doubt that flavor changing processes exist for fast partons
coupling to a medium. Indeed, flavor changing channels have been included
in computations of medium-modified fragmentation functions
\cite{Schafer:2007xh}, and quark-gluon conversions of the leading jet
particle are also implemented in the AMY formalism \cite{AMY:02}.
Nevertheless, flavor changing processes are often neglected in
phenomenological studies.

This situation led to the wide-spread opinion that quark and gluon jets
are well-defined concepts in a medium and that they ought to exhibit a relative
suppression by a color factor $9/4$.
This claim might have some validity in the case that the mean-free path
of the leading particle is of the order of (or larger than) the size of
the medium, $\lambda \approx L$. Energy loss is then dominated by emission of
only one gluon. In general, however, the situation is different. We want
to emphasize that the flavor of a leading parton might only be well-defined
locally and that it can change along the trajectory.
Recently, Sapeta and Wiedemann argued that changes in jet chemistry could
also arise from the increased multiplicity in medium-modified jets
\cite{Sapeta:2007ad}. In a model using enhanced parton branching they
find effects similar to those discussed by us here.

Nevertheless, most studies of jets in nuclear matter still focus solely on the
\emph{kinematical} effects, based on quenching of longitudinal momentum
and broadening of transverse momentum of the leading parton. E.g.,
observables like $R_{AA}$ and $v_2$ are sensitive to various integrals
over the differential energy loss $dE/dx$ and ultimately measure the momentum
transfer per path length, i.e.\ $\hat q$. We can summarize this as the
effect of jets coupling to the \emph{thermal} properties of the medium.
We like to advocate a second, complementary approach which looks at
\emph{flavor} effects from jets coupling to the \emph{chemical} properties
of the medium. Measurements of identified hadrons at high-$p_T$ can
constrain the rate of conversions. With sufficient experimental
sensitivity this would lead to estimates for the mean free path
$\lambda$ of the jet in the medium, complementary to measurements
of $\hat q$. We expect such measurements to provide additional
stringent tests for the validity of any model for the jet-medium coupling.
Of course, with any novel observation different explanations have to
be taken into account.

This work is organized as follows. In the next section we will discuss
all relevant conversion channels on a qualitative level and assess their
relative importance. We focus on jets
interacting with a deconfined quark gluon plasma in nuclear collisions,
but our arguments can be easily applied to cold or hot hadronic matter as well.
We then proceed to present numerical results for RHIC and LHC in a consistent model based on rate equations for conversions and a Fokker-Planck equation for
energy loss, using perturbative leading-order (LO) cross sections.
The formalism is introduced in Sec.\ \ref{propagate} while a discussion of
the results can be found in Secs.\ IV and V.
We focus on light and strange quarks, and real photons. Conversions including
heavy quarks will not affect the abundance of light species and photons
measurably. Therefore, the effect on charm and bottom spectra will be
addressed in
a separate publication \cite{weiliu3}. We will also introduce some novel double-ratio
observables that might be particularly sensitive to conversion processes.
Finally, a summary and discussion are presented in Sec.\ \ref{summary}.

\section{Flavor Conversions in Quark Gluon Plasma}

Let us start by discussing possible flavor conversion processes in a
leading order perturbative approach. Annihilation and pair creation
processes (i) $q+\bar q \leftrightarrow g+g$ can lead to a conversion of quarks into gluons and vice versa.
In the following we agree that the first parton mentioned on either side of an
arrow $\leftrightarrow$ or $\to$ is the leading jet parton in the sense that
it is the one with the larger momentum with respect to the local rest frame of
the medium among all other partons on the same side of the arrow. $q$ can
be a quark or antiquark and $\bar q$ is its antiparticle.

Photons can be easily created through (ii) $q+\bar q \to \gamma + g$ and (iii)
$q +\bar q \to \gamma+\gamma$. Note that we can neglect the opposite processes
and generally all processes with photons in the initial state for obvious
reasons. It is also safe to ignore $q+\bar q \to g + \gamma$ which in our
notation contributes to the yield of gluon jets, but is much suppressed
compared to the annihilation contribution (i).
On the other hand, flavor changing Compton processes
(iv) $q+g \leftrightarrow g +q$ are as important as annihilation and pair
creation. Compton processes also contribute to photon production via
(v) $q+g \to \gamma+q$.

Let us comment on the fate of heavy quarks. Heavy quark annihilation is
negligible at realistic temperatures, but the opposite process (vi)
$g+g \to Q + \bar Q$ might be an important source of excess heavy quarks.
The Compton process for heavy quarks (vii) $Q+g \to g+ Q$ is
interesting as well since it can effectively accelerate heavy quarks from
the medium or decelerate existing fast heavy quarks. We will
report on the quantitative effects of
processes involving heavy quarks in a forthcoming publication \cite{weiliu3}.

Some of the conversion channels (i) to (v) have been investigated in the
past. In \cite{weiliu1} the authors discussed the conversion of quark
into gluon jets and vice versa through annihilation and Compton channels
(i) and (iv), respectively. In the absence of flavor conversions the relative
suppression of gluon relative to quark jets is expected to be 9/4.
Conversions reduce the difference between the nuclear modification factors
for high-$p_T$ quarks and gluons. Assuming gluon dominated proton fragmentation
functions, as given in modern parameterizations \cite{AKK:05}, this leads to
an enhancement of the $p/\pi^+$ and ${\bar p}/\pi^-$ ratios at high transverse momentum. This is
supported by recent data from the STAR Collaboration at RHIC \cite{adams1,
adams} which shows rather large $p/\pi^+$ ratios in Au+Au at high $p_T$.
Models without conversion predict consistently lower $p/\pi^+$ and
$\bar p/\pi^-$ ratios \cite{wang1}.

Interestingly, in \cite{weiliu1} the required conversion rates were found to
be much larger than those given by the leading order perturbative
calculation. An enhancement factor ($K\approx 4$) was necessary to describe the
data. Although there are some uncertainties from the fragmentation functions
this either points to large higher order corrections \cite{weiliu2} or
to a non-perturbative mechanism favored in a strongly coupled QGP scenario
\cite{whitepaper:05}.

It was first pointed out in \cite{fries1} that annihilation and Compton
processes can also lead to the conversion of a quark or gluon jet into a
photon. Perturbative estimates of this process at leading order indicate a
sizable contribution to the total direct photon yield at intermediate $p_T$
between 3 and 6 GeV/$c$ at RHIC energies \cite{FMS:05,simon,GAFS:04,
Turbide:2007mi}. The same was found for the production of virtual photons
and lepton pairs \cite{SGF:02,Turbide:2007mi}.
Measurements of single inclusive direct photon production from the
PHENIX experiment \cite{phenix:photons} are not conclusive so far. The data
has been described equally well within error bars both by calculations
taking into account only prompt hard photons and thermal photons,
and by calculations adding jet-photon conversions as well
\cite{Isobe:07}.

Recently, it was also noticed that jet-photon
conversions lead to photons with an azimuthal asymmetry $v_2$ which is
negative w.r.t.\ the $v_2$ of hadrons and other photon sources
\cite{simon1}. In Ref.\ \cite{simon1} it was estimated that the sum
of all direct photon sources leads to small and negative direct photon
$v_2$ between 3 and 5\% at intermediate $p_T$. However, later it was noted
that the result depends crucially on details of the fireball simulation
and might be much closer to zero \cite{Turbide:2007mi}. First experimental
results are compatible with zero with sizable uncertainties
\cite{phenix:05v2,phenix:07v2}.

The current experimental situation can be summarized by noting that
both for conversions into photons and for quark-gluon conversions no
final conclusion has been reached about a direct observation of these
channels. This should change in the future with extended hadron
identification capabilities at RHIC and increased statistics for
direct photons. Of course, the experimental confirmation of one
conversion signal would be a strong indication that conversion in
general, as advocated here, is an important process.

In the remainder of the paper we present a calculation of conversion
channels (i),(ii),(vi) and (v) in one consistent framework. We are going
to compute
nuclear modification factors $R_{AA}$ for pions, kaons and direct photons,
and the elliptic flow $v_2$ of direct photons. This leads to a
direct quantitative connection between the conversions into photons
\cite{fries1} and conversions into gluons \cite{weiliu1}.

Furthermore, we suggest the relative yield of strange hadrons at high $p_T$
as a new signature for jet medium coupling. The ratio of $s$ quarks to
the sum of $u$ and $d$ quarks,
\begin{equation}
  w = \frac{s}{u+d}
\end{equation}
is about 5\% for the initial leading jet particle at RHIC energies at a
typical $p_T$ of about 10 GeV/$c$. It can be easily traced back to the
small fraction of strange quarks in the incident nucleons. A rough estimate
with CTEQ5M parton distributions \cite{cteq5:99}, assuming dominance of Compton
channels in the initial hard scattering would give
\begin{align}
  w_\text{jet}(p_T = 10 \text{ GeV/}c) &\approx w_\text{pdf}(x\approx 0.1,
  Q\approx 10 \text{ GeV}) \nonumber \\
  &\approx \frac{0.07}{1.1} \approx 6.4 \% \, ,
\end{align}
consistent with the result from the full calculation.
On the other hand, in a chemically equilibrated quark gluon plasma at a
given temperature $T$ the ratio is
\begin{equation}
  w_\text{ce}(T) \approx \frac{m_s^2}{4T^2} K_2(m_s/T) \, .
\end{equation}
assuming massless $u$ and $d$ quarks. For a strange quark mass of $m_s = 100$
MeV the ratio is almost 1/2 even at $T_c \approx 180$ MeV:
$w_\text{ce}(T_c) = 0.47$. The difference between $w_\text{jet}$ and
$w_\text{ce}$ is rather large and we have to expect that the relative
abundance of strange quark jets will rise with time.

For an infinite medium the particle ratios of the jets
would equilibrate to that of the medium (the same will happen to their
momentum distribution, making them indistinguishable from the
medium). For a finite path length $L$, the rate of
equilibration should be a good measure of the strength of the coupling
to the medium. In particular, it should give a good estimate of
the mean free path $\lambda$ between flavor changing scatterings.
The approach to equilibrium will be determined by the ratio $\lambda/L$.

The final parton abundances, once a jet leaves the medium, will translate
into hadron abundances as given by the fragmentation process. Our poor
knowledge of fragmentation functions limits our ability to make precise
predictions for absolute hadron ratios. Hence, we propose to use double
ratios of abundances measured in $A+A$ and $p+p$ collisions. The first ratio
we are going to look at is that for direct photons relative to pions
\begin{equation}
  r_{\gamma/\pi^+} = \frac{(\gamma/\pi^+)_{AA}}{(\gamma/\pi^+)_{pp}}.
\end{equation}
Note that formally this is the same as $R_{AA}^\gamma/R_{AA}^{\pi}$.
In such double ratios uncertainties from the determination of the number
of $N+N$ collisions (needed for $R_{AA}$) cancel. On the theoretical side
the large uncertainties from fragmentation functions tend to cancel between
the $A+A$ and the $p+p$ contribution. On the experimental side, we would
hope that the systematic errors for a direct extraction of double ratios
are smaller than for an a posteriori reconstruction from single ratios.
We will also take a fresh look at the relative abundances of protons and
pions
\begin{equation}
   r_{p/\pi^+} = \frac{(p/\pi^+)_{AA}}{(p/\pi^+)_{pp}}.
\end{equation}

Generally, in $A+A$ collisions we expect the relative abundances of flavors
to tend toward equilibrium values. Therefore, we expect an increase of
photons compared to quark and gluons and we expect an increase of gluons
relative to quarks \cite{weiliu1}. Under the assumption that protons
are dominated by fragmentation from gluons we expect the $p/\pi^+$ ratio
to increase in $A+A$ collisions, while $\gamma/\pi^+$ will show
an even steeper increase. A similar increase should also be observed in the
relative abundance of strange hadrons as high-$p_T$ strange quarks tend
toward equilibrium.
In the remainder of the paper we will try to verify our claims numerically.

\section{Jet Propagation in an Expanding QGP}\label{propagate}

Our treatment of a fast quark or gluon in a QGP medium has
already been discussed in Ref.\ \cite{weiliu,weiliu1}. The behavior is
governed by coupled Fokker-Planck equations and rate equations
\begin{eqnarray}\label{fpeq}
\frac{d\langle p^a_T \rangle}{d\tau} &=& -\langle \gamma_a(p_T,T) p^a_T
\rangle \approx -\gamma_a(\langle p^a_T\rangle,T))\langle p^a_T\rangle \, ,
\\
\frac{dN^a}{d\tau} &=& -\sum_{b}\Gamma^{a\to b}(p_T,T)N^a+\sum_{c}\Gamma^{c\to a}(p_T,T)N^c,\nonumber
\end{eqnarray}
with letters $a,b,c$ denoting quark, antiquark, gluon, or photon.
We have dropped the diffusion term and also neglected the
dispersion of jet momentum, i.e.\ $\langle p^2_T\rangle = \langle p_T\rangle^2$
since we want to focus on high momentum jets. The drag coefficient $\gamma_a$
is obtained from Ref.\ \cite{weiliu1}, including all elastic processes
between quarks and gluons.
To mimic the contributions from higher order radiative processes, we introduce an enhancement factor $K$.
The conversion widths $\Gamma$ for quark to gluon and vice versa (channels
(i) and (iv)) are extracted from Ref.\ \cite{weiliu1} as well, multiplied by the
same enhancement factor $K$ to account for higher order corrections. The widths
for the quark to photon processes (ii) and (v) at leading order are obtained
from Ref.\ \cite{fries1}.

For a quark or gluon jet moving through the QGP, the rate of change
for its mean transverse momentum $\langle p_T\rangle$ can be obtained from the
Fokker-Planck equation in (\ref{fpeq}). A quark or gluon can be
converted to a gluon, quark or photon with a rate given by the corresponding
collisional width. These effects are modeled by introducing a large number of
test quark and gluon jets that are distributed in the transverse plane
according to the underlying binary nucleon-nucleon collisions. Their
transverse momentum distribution is taken to be uniform in azimuthal angle
$\phi$. A test jet of a given kind with certain transverse momentum is
assigned a probability that is proportional to the corresponding jet momentum
spectrum with the proportionality constant determined by requiring that the
sum of the probabilities for all test jets of this kind is equal to their
total number.
More details can be found in Ref.\ \cite{weiliu1}.

Let us quickly discuss the possible lessons that can be learned by looking
at a simplified situation. We assume a rare jet flavor which can be produced
from a plentiful reservoir of a second jet flavor via a single
channel with width $\Gamma\sim 1/\lambda$. As long as the first
flavor stays rare the back-reaction is negligible.
A real world example close to this hypothetical situation could be photons
produced from light quark jets at RHIC.
Then the rate to produce an additional excess of the rare flavor is
\begin{equation}
  \frac{dN_\text{rare}}{d\tau} \approx \frac{1}{\lambda} N_\text{res}
\end{equation}
Let us further assume that $N_\text{res}$ is roughly constant during the
lifetime of jets in the system (as for the real world example: up to 30\%
of quark jets at RHIC leak to gluon jets \cite{weiliu1}).
Then we arrive at a simple formula connecting yields and the mean
free path for the single channel.
\begin{equation}
  \frac{\lambda}{L} \approx \frac{N_\text{res}}{N_\text{rare,excess}}.
\end{equation}

Obviously, realistic systems are more complicated with multiple channels,
changing rates $\Gamma$ due to the temperature evolution of the system,
and a whole range of possible path lengths $L$.
Therefore, in the following section we will apply our test particle Monte Carlo simulation to study the propagation of jets in an expanding QGP fireball.

\section{Results for Spectra , $R_{AA}$ and $v_2$ at RHIC} \label{nucl-mod}

\begin{figure}[t]
\centerline{
\includegraphics[width=3.0in,height=3.0in,angle=-90]{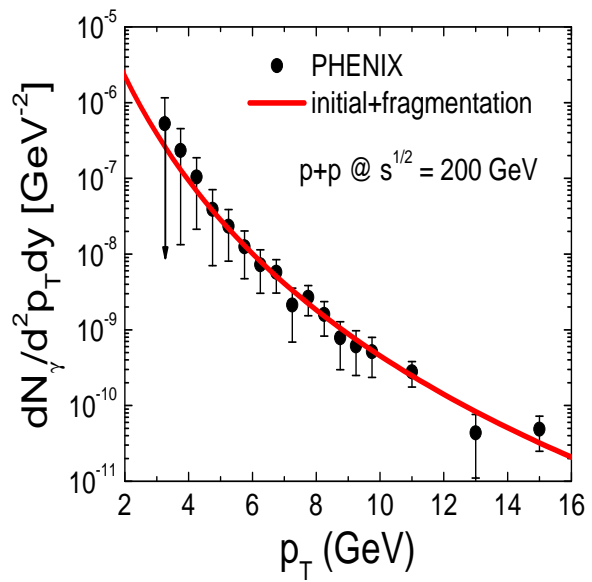}}
\caption{(Color online)Photon spectrum in $p+p$ collisions at
  $\sqrt{s_{NN}}=200$ GeV as function of transverse momentum.
  Data are taken from the PHENIX Collaboration \cite{adler2}.}
\label{spectra_pp}
\end{figure}

In $p+p$ collisions, direct photons come from initial hard
collisions and vacuum bremsstrahlung off jets.
Bremsstrahlung photons can be obtained by folding the QCD
with photon fragmentation functions
$D_{\gamma/(q,g)} (z,Q^2)$ from Ref.\ \cite{owens}. The spectra of quark and gluon jets are taken from
\cite{weiliu2} which fit both pion and proton spectra measured in $p+p$
collisions from STAR, while the prompt initial photons are calculated
at leading order according to \cite{owens} with a $K$ factor of 1.5.
The resulting $p_T$ spectrum of photons in $p+p$
collisions at $\sqrt{s_{NN}}$=200 GeV is shown in Fig.\ \ref{spectra_pp}, and it
is found to be in good agreement with data from PHENIX \cite{adler2} at
transverse momenta above 4 GeV/$c$. This provides a well controlled baseline
for calculations of the nuclear modification factor $R^\gamma_{AA}$.

\begin{figure*}[ht]
\includegraphics[width=3.0in,height=3.0in,angle=-90]{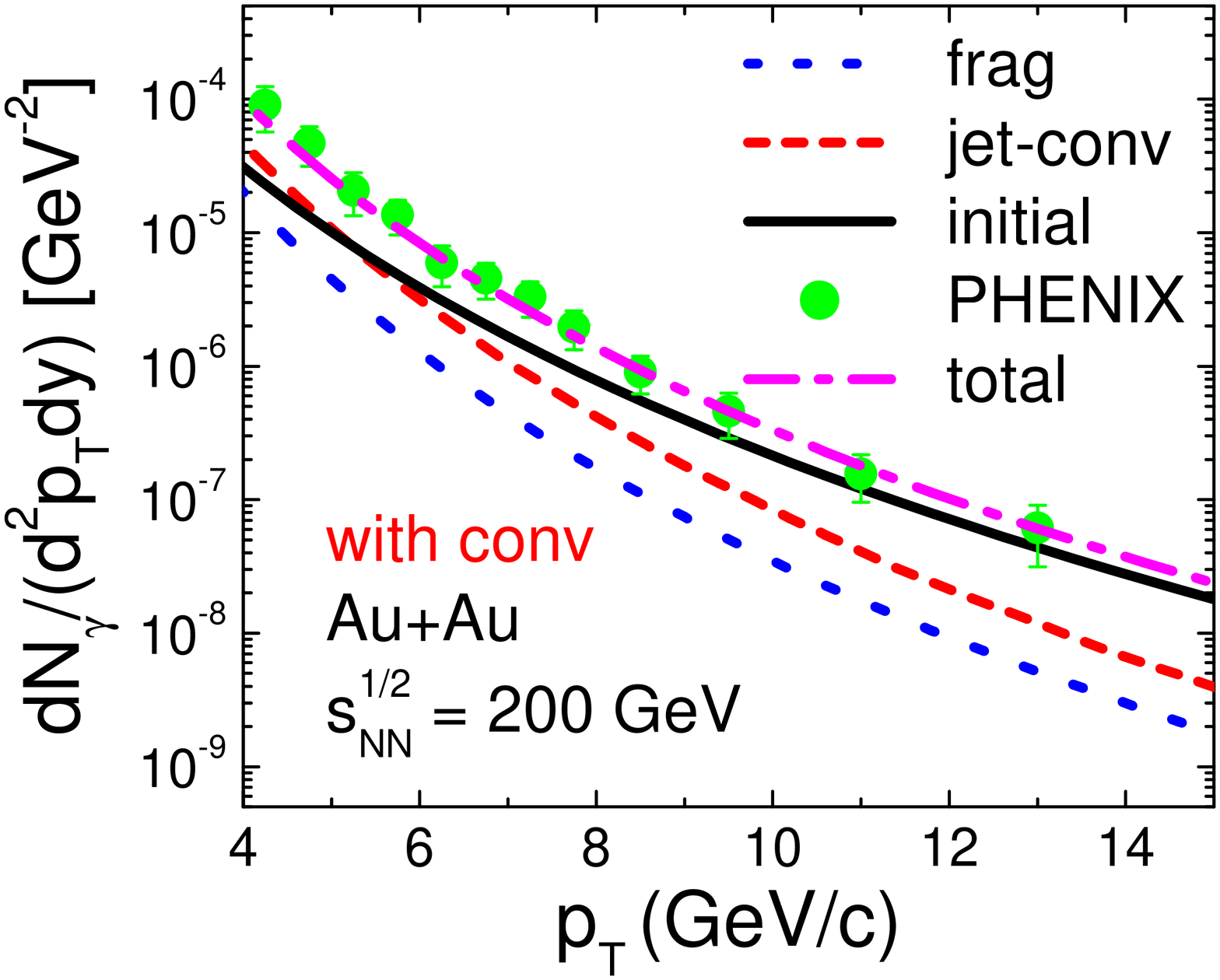}
\hspace{0.2cm}
\includegraphics[width=3.0in,height=3.0in,angle=-90]{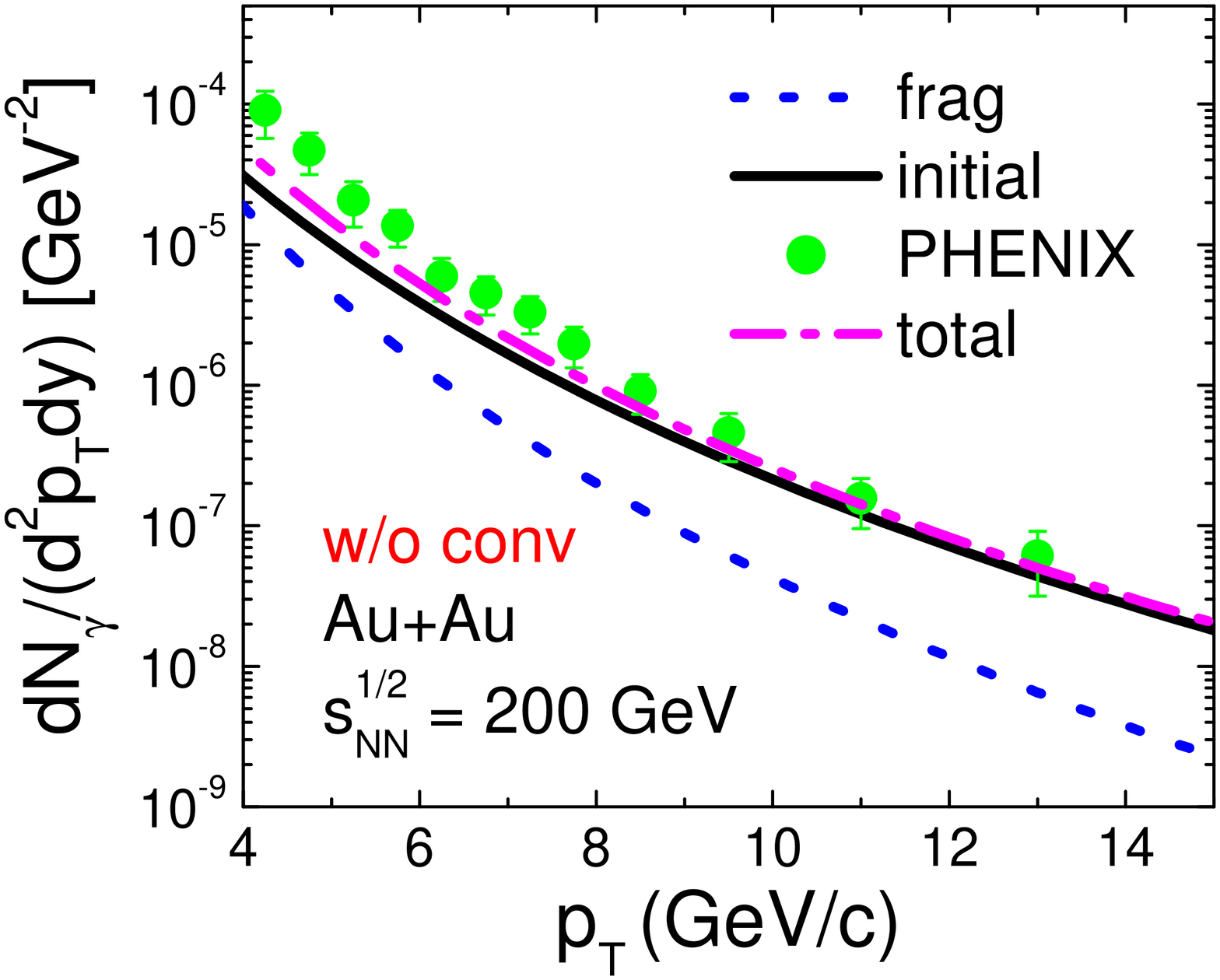}
\caption{(Color online) Photon spectra in central Au+Au collisions at
$\sqrt{s_{NN}}=200$ GeV as functions of transverse momentum with (left panel)
and without conversions (right panel).  Direct photons are from initial hard
collisions (Solid line), final state jet fragmentation (Dotted line), and
jet-photon conversions (dashed line). Data are taken from PHENIX \cite{adler4}.}
\label{spectra}
\end{figure*}

The yield of additional direct photons in Au+Au collisions from jets in the
medium is computed by
applying the test particle Monte Carlo method discussed above. For the
dynamics of the fireball in central collisions we assume that it evolves
boost invariantly in the longitudinal direction, but with an accelerated
transverse expansion. The parameters for the fireball formed in Au+Au
collisions at $\sqrt{s_{NN}}=$ 200 GeV are taken from Refs.\
\cite{weiliu, weiliu1, chen} where one of us studied energy loss of
heavy quark, and the $p/\pi^+$ and $\bar p/\pi^-$ ratios due to jet flavor conversions, as well as production of pentaquark baryons.
The volume expands in the proper time $\tau$ according to
$V(\tau)=\pi R(\tau)^2\tau$, where $R(\tau) = R_0 + a(\tau-\tau_0)^2/2$ is the
transverse radius with an initial value $R_0$=7 fm/$c$, $\tau_0$=0.6 fm/$c$ is
the QGP formation time, and $a=0.1 \,c^2$/fm is the transverse
acceleration. Starting with an initial temperature $T_0=350$ MeV, the time
dependence of the temperature is obtained from entropy conservation, leading
to the critical temperature $T_c=175$ MeV at proper time $\tau_c=5.0$ fm/$c$.

We show all results with two extreme assumptions for the $K$-factor of
the jet-plasma interactions. We explore the scenario $K=0$ which simply
switches off all flavor conversions. We also show computations with
$K=4$. This rather large value is likely to establish an
upper bound for the effect, but it is also compatible with the previous
work in \cite{weiliu1}. Of course, taking such a large value would imply
that a leading order perturbative treatment is not sufficient to begin
with. Note that we keep a separate value of $K = 1$ for conversions
with a photon in the final state. We have fixed this value with the available
data on direct photons.

The resulting spectra of direct photons in Au+Au collisions at
$\sqrt{s_{NN}}$ = 200 GeV are plotted in Fig.\ \ref{spectra} as functions of
transverse momentum. The left panel shows the scenario with jet flavor and jet-photon conversions included,
while the right panel neglects conversions. The results show that direct photons from jet-photon conversions (dashed line in the left panel)
make a sizable contribution below $p_T= 6$ GeV/$c$ at RHIC, in accordance with
previous results \cite{fries1}. Jet flavor conversions reduce the number of
high momentum quark jets by 30\% \cite{weiliu1}. Therefore, conversions lead to
a decreased production of direct photons from final state jet fragmentation
(dotted lines in both left panel and right panel) by roughly the same amount.

\begin{figure*}[tb]
\includegraphics[width=2.3in,height=2.2in,angle=-90]{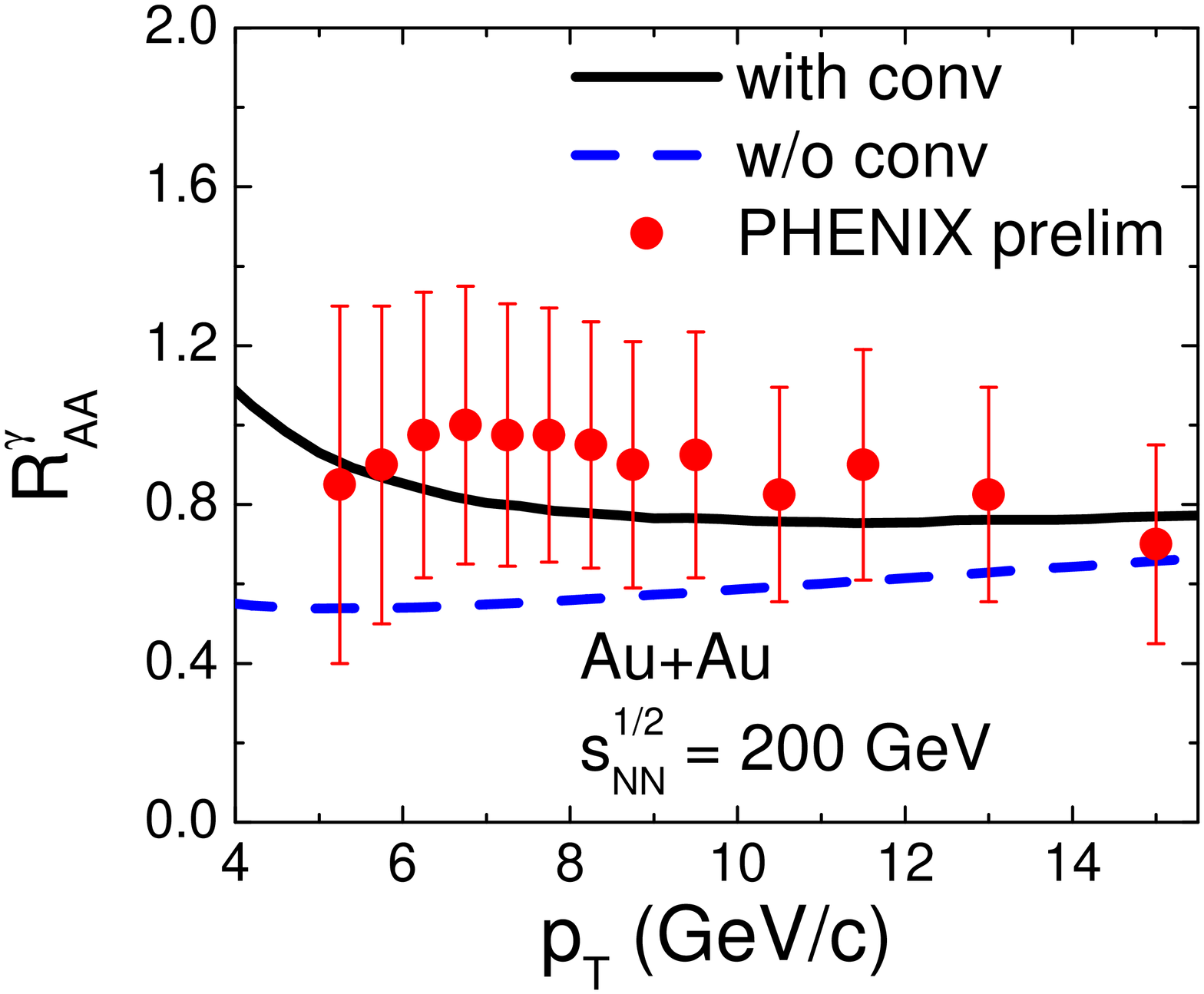}
\hspace{0.2cm}
\includegraphics[width=2.3in,height=2.2in,angle=-90]{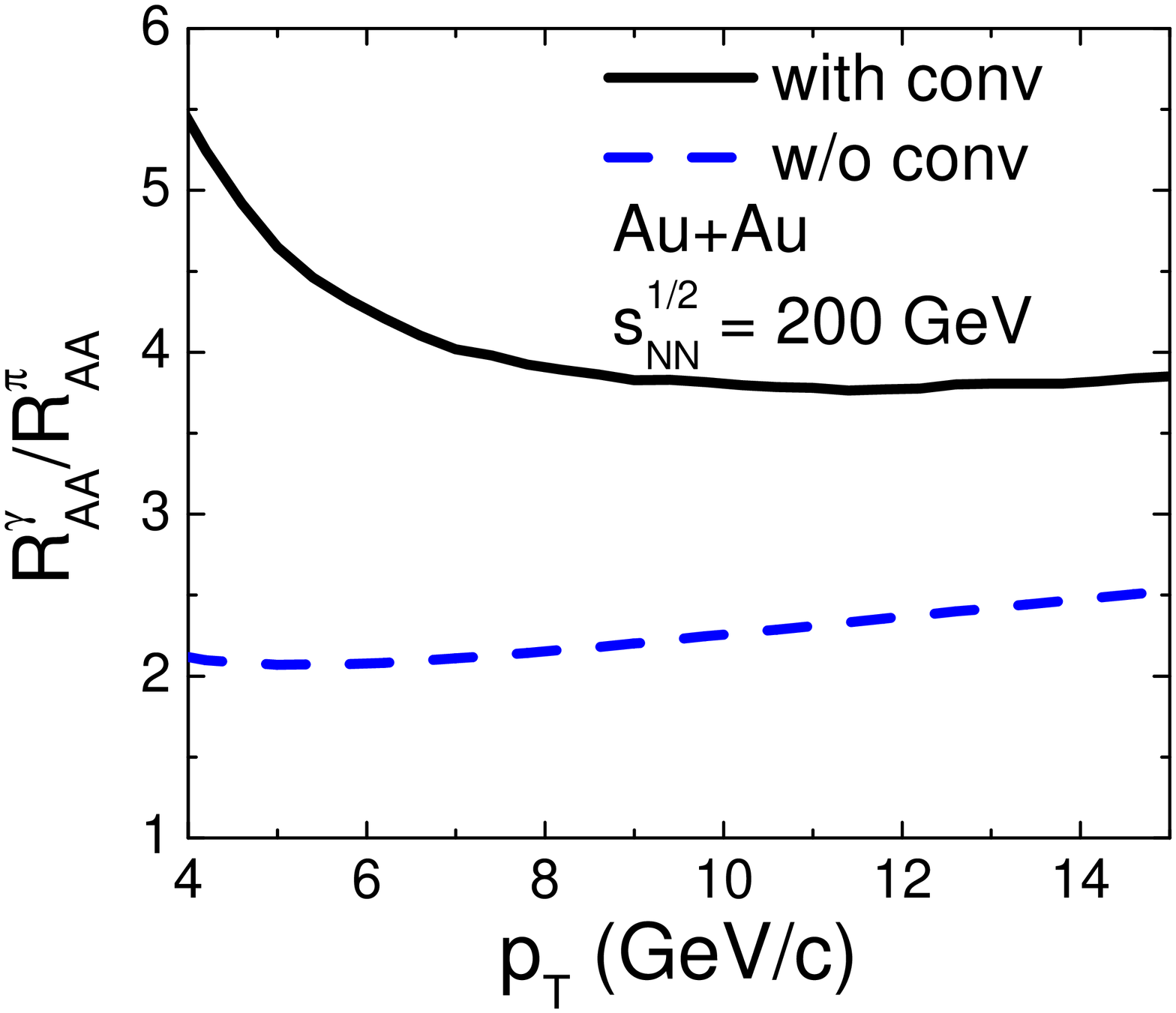}
\hspace{0.2cm}
\includegraphics[width=2.3in,height=2.2in,angle=-90]{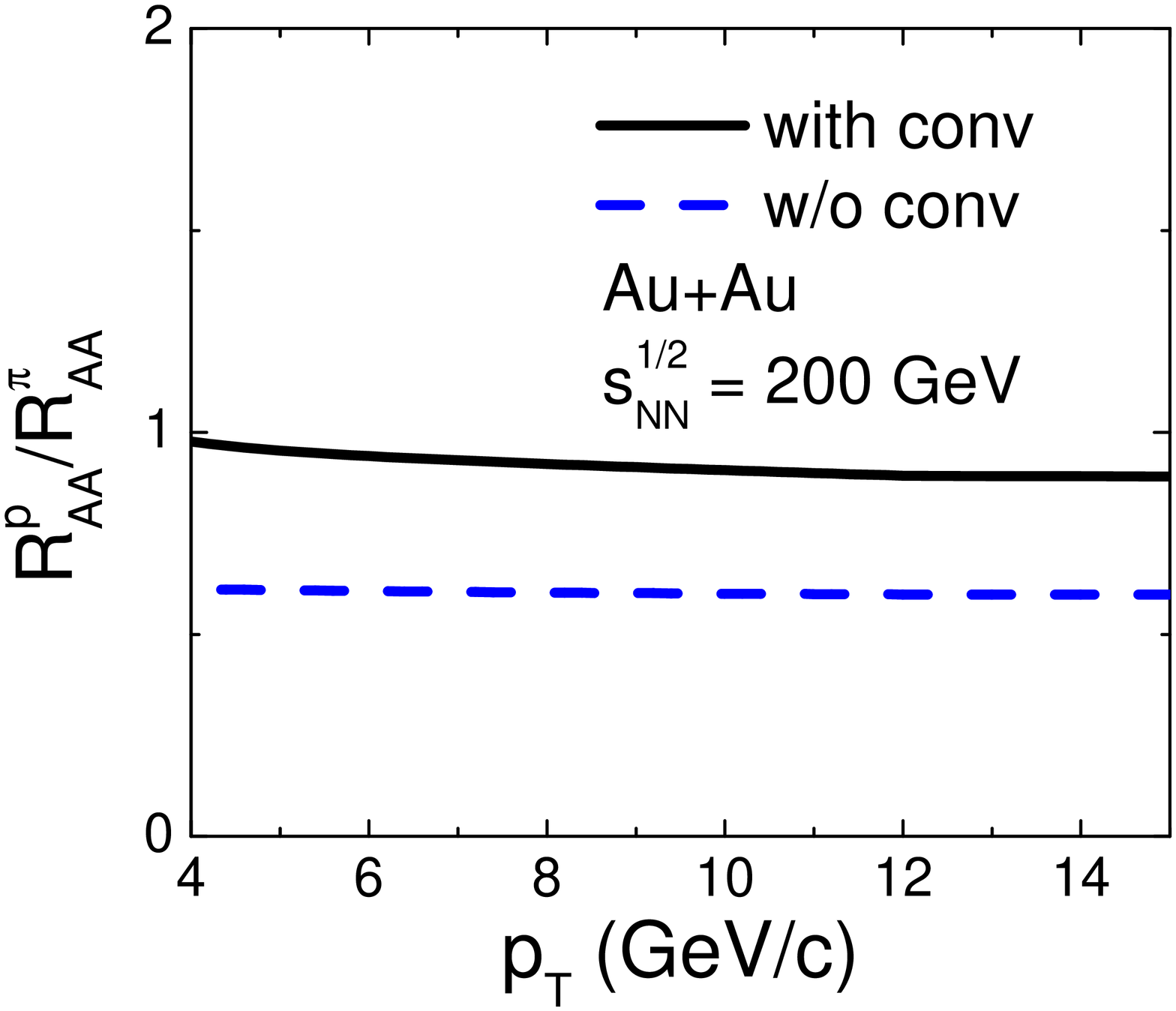}
\caption{(Color online) Nuclear modification factor $R^\gamma_{AA}$
of direct photons (left panel) and the double ratios of $r_{\gamma/\pi^+} = R^\gamma_{AA} / R^{\pi^+}_{AA}$ (middle panel) and $r_{p/\pi^+} =
R^p_{AA}/R^{\pi^+}_{AA}$ in central Au+Au collisions at $\sqrt{s_{NN}}=200$ GeV as functions of transverse momentum $p_T$. Results are with (solid line) or without (dotted line) conversions.}
\label{raa}
\end{figure*}

The nuclear modification factor $R^\gamma_{AA}$ for the total direct
photon spectrum is shown in the left panel of Fig.\ \ref{raa} (the number of binary collisions is taken to be $\approx$ 960 for the most cental bin).
The result with all conversion turned off ($K=0$) is plotted with a dotted line
while that with jet conversions included is indicated
by the solid line. We find a noticeable difference between both cases. In our
calculation the result with conversions is more compatible with the data
from PHENIX. Note that the difference between both scenarios
decreases with increasing transverse momentum because prompt hard photons from
initial scatterings, unchanged by the nuclear environment, dominate at higher
$p_T$.

As we discussed above, it might be interesting to look at double ratios of nuclear modification factors to extract jet conversion effects.
In the middle panel of Fig.\ \ref{raa} we present our expectation for
$r_{\gamma/\pi^+}= R^\gamma_{AA}/R^{\pi^+}_{AA}$ at intermediate and high
$p_T$. Note that no recombination effects are taken into account in this study
\cite{reco}, so that deviations in quantities involving hadrons can be
expected below $p_T \approx 6$ GeV/$c$. We find that conversion processes lead
to a large signal in $r_{\gamma/\pi^+}$. Our results with and without conversion
differ by a factor 2. We also show the double ratio $r_{p/\pi^+} =
R^p_{AA}/R^{\pi^+}_{AA}$ in the right panel of Fig.\ \ref{raa}. Again, we find a
large separation between the extreme cases with and without conversions,
opening the door for a possible measurement at RHIC.


\begin{figure*}[t]
\centerline{
\includegraphics[width=2.5in,height=2.5in,angle=0]{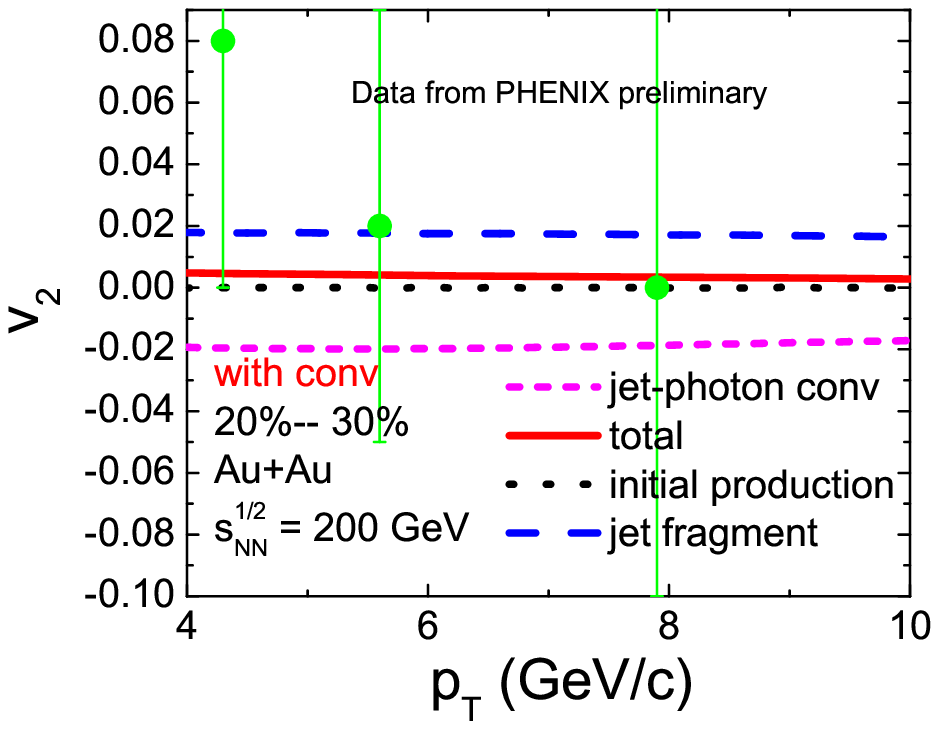}
\hspace{-0.2cm}
\includegraphics[width=2.5in,height=2.5in,angle=0]{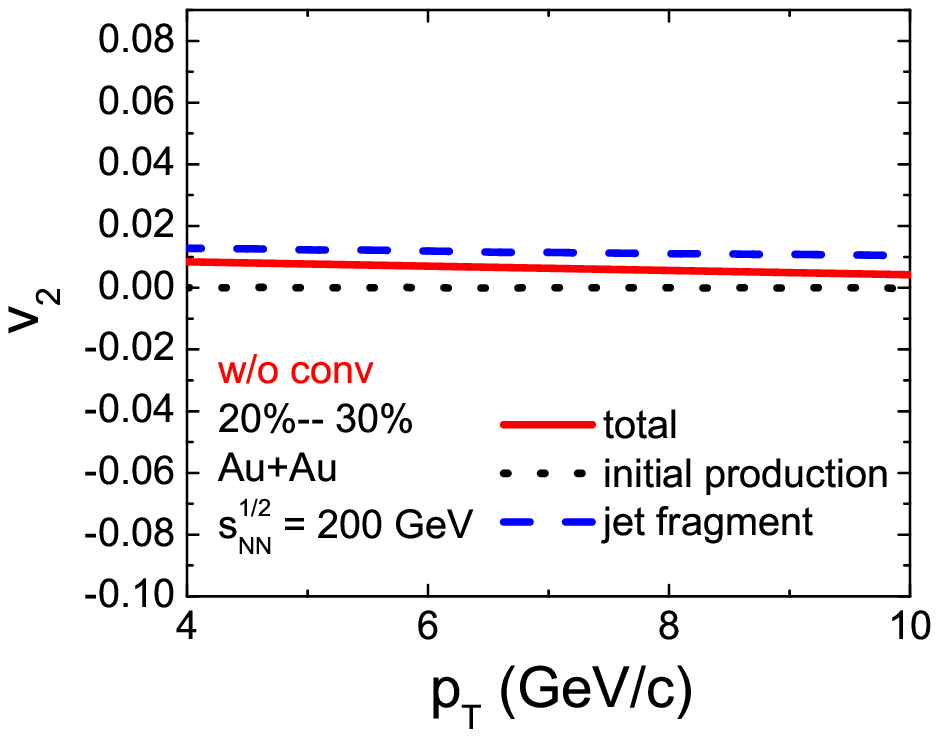}}
\caption{(Color online) Elliptic flow $v^\gamma_2$ of direct photons in
  peripheral Au+Au collisions with centrality of $20 -30\% $ at
  $\sqrt{s_{NN}}=200$ GeV as functions of transverse momentum for two
  different scenarios of with (left panel) or without conversions (right
  panel), respectively. Data is taken from the PHENIX collaboration.
\cite{phenix:07v2}}
\label{v_2}
\end{figure*}

\begin{figure*}[t]
\includegraphics[width=3.0in,height=3.5in,angle=-90]{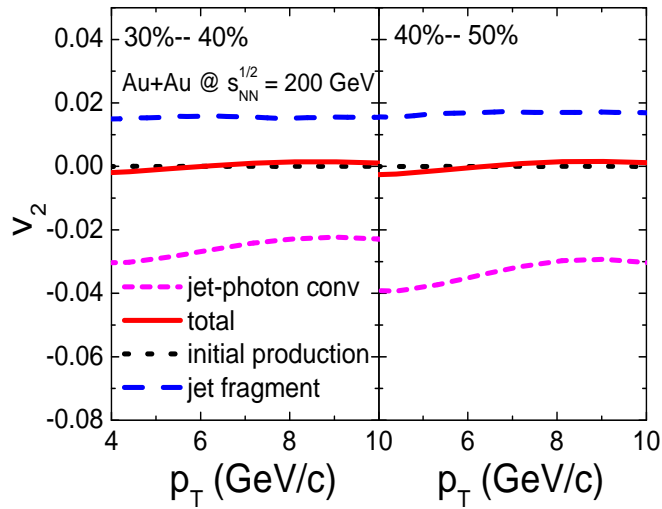}
\caption{(Color online) Elliptic flow $v^\gamma_2$ of direct photons in
  peripheral Au+Au collisions with centrality of $30 -40\% $ (left) and $40 - 50\% $ (right) at $\sqrt{s_{NN}}=200$ GeV as a function of transverse momentum.}
\label{v_2_diff_acce}
\end{figure*}

An azimuthal anisotropy is believed to be generated in a partonic stage in semi-central and peripheral heavy ion collisions when the fireball has
the maximum geometric asymmetry. This anisotropy is studied in terms of Fourier coefficients $v_2$ defined from particle yield $dN/(p_Tdp_Td\phi)$ as
\begin{eqnarray}
  \frac{dN}{p_Tdp_Td\phi} = \frac{dN}{2\pi p_Tdp_T}\left[1+ 2v_2(p_T)\cos
  (2\phi)+ \mathcal{O}(4\phi)\right]
\end{eqnarray}
where the angle $\phi$ is defined with respect to the reaction plane.
The coefficient $v_2$ can be extracted from the transverse momentum $p_T = (p_x,p_y)$ of measured particles via
\begin{eqnarray}\label{eq-v2}
v_2 = \left\langle \frac{p^2_x-p^2_y}{p^2_x+p^2_y}\right\rangle,
\end{eqnarray}
where the direction of the $x$-axis is in the reaction plane with
$\phi$ = 0 and the direction of the $y$-axis is out of the reaction plane.
Here, $\langle \cdots\rangle$ means averaging over all measured particles
for a given species.

The parameters of the expanding fireball formed in peripheral collisions
are extracted from information about centrality, number of participants, and
number of binary collisions in Au+Au collisions at $\sqrt{s_{NN}}$= 200 GeV
given in \cite{adler3}. For example, in collisions in the centrality bin 20--30\%, the fireball forms an almond-shaped cylinder with the initial length of the long transverse axis being $l^0_L$ = 6.1 fm and that of the short transverse axis being $l^0_S$ = 3.5 fm.
The acceleration in the direction of the long axis is $a_L$ = 0.1 $c^2$/fm and that in the direction of the short axis is $a_S=l_L/l_S a_L$. We also
assume the fireball is boost-invariant and
the temperature dependence is obtained from entropy conservation with an
initial temperature $T_0$ = 340 MeV at proper time $\tau_0$ = 0.6 fm/$c$ and critical temperature $T_c$ = 175 MeV.

\begin{figure*}[tb]
\includegraphics[width=3.0in,height=3.0in,angle=-90]{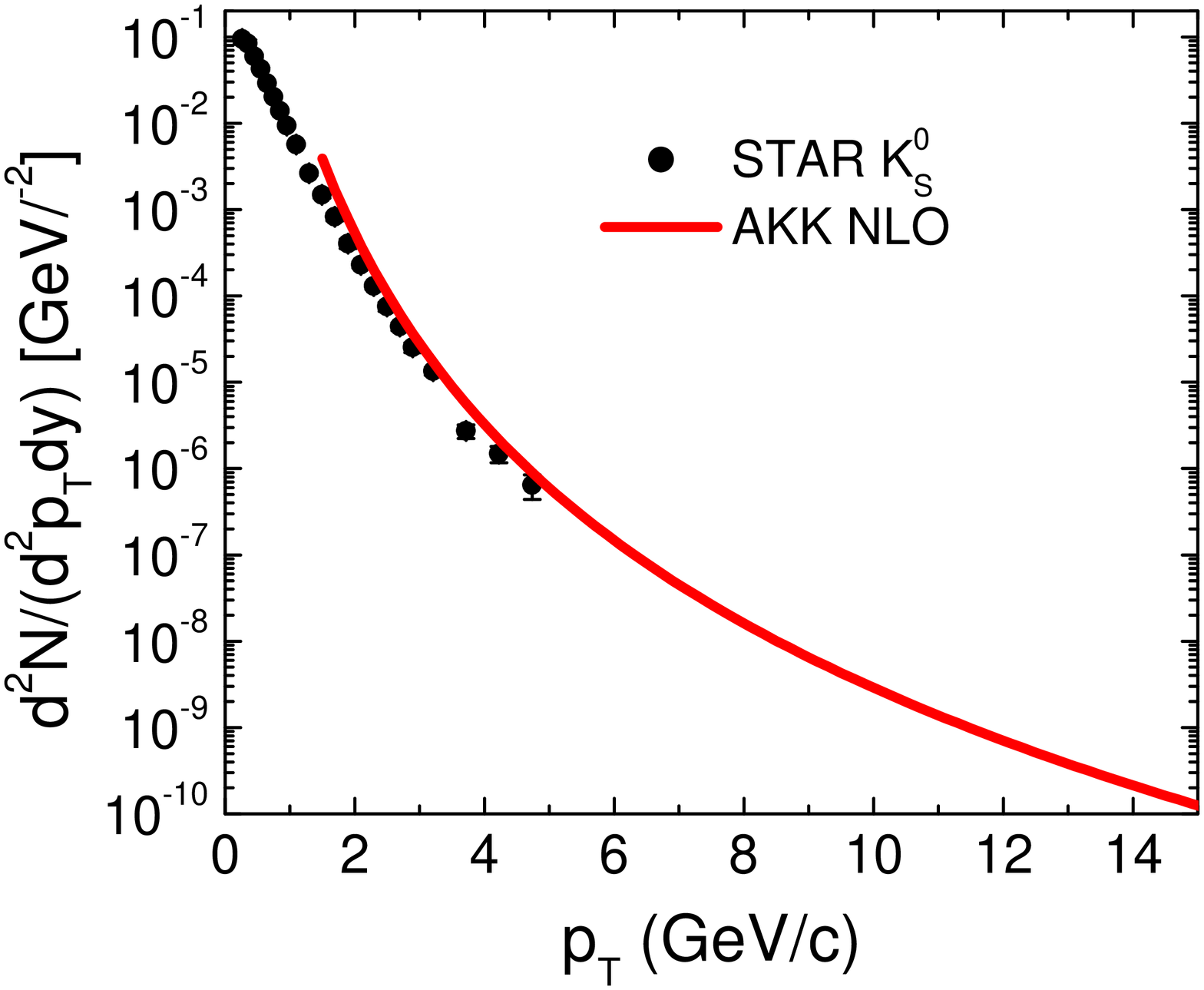}
\includegraphics[width=3.0in,height=3.0in,angle=-90]{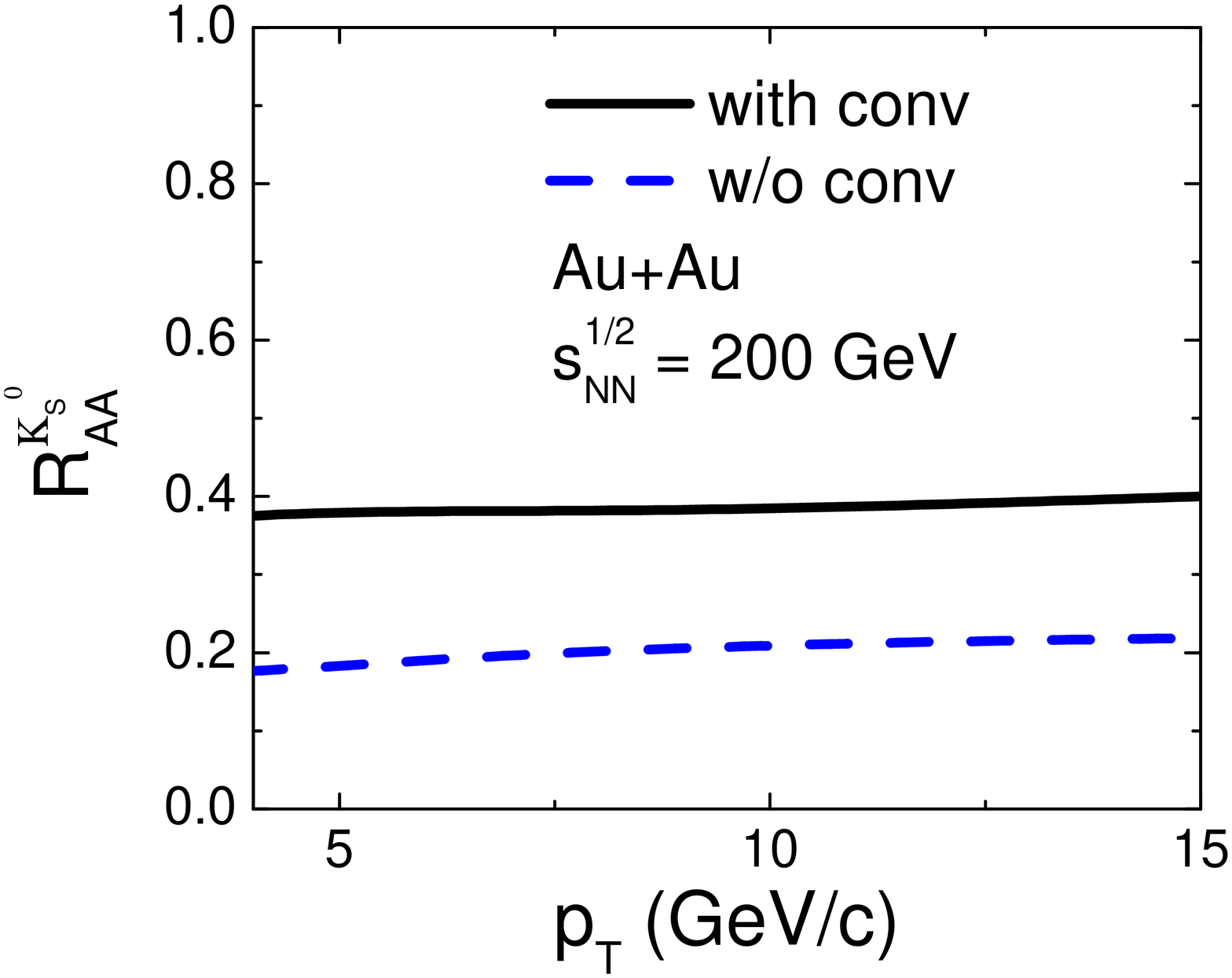}
\caption{(Color online) $K^0_S$ (left panel) spectra from quark and gluon jet
  fragmentation via AKK fragmentation function \cite{AKK:05} in $p+p$
  collisions at $\sqrt{s_{NN}}=$200 GeV. Data are from the STAR Collaboration
  \cite{abelev}. Nuclear modification factor $R_{AA}$ (right panel) for
  $K^0_S$ in Au+Au collisions at same center-of-mass energy as a function of transverse momentum.}
\label{raa_k0}
\end{figure*}

The obtained elliptic flow $v^\gamma_2$ of direct photons is shown in Fig.\
\ref{v_2} as a function of $p_T$. We plot the results with conversions in the
left panel, while those in the right panel have $K=0$. The elliptic flow of
photons from jet-photon conversions is negative with a magnitude of about 2\%,
because the production of direct photons in the direction out of the reaction
plane is favored by the longer propagation path of jets through the medium
\cite{simon1}. This makes it more likely that a rare jet-photon conversion
occurs. $v^\gamma_2$ of photons from final state jet fragmentation is
positive, following the pattern of $v_2$ for hadrons at large $p_T$. Adding
all contributions together leads to a positive total elliptic flow
$v^\gamma_2$ of direct photons. When jet conversions are switched off, the elliptic flow
$v^\gamma_2$ of direct photons from final state jet fragmentation decreases
because of the larger number of quark jets leaving the fireball. The effects
of missing negative conversion $v_2$ and increased positive bremsstrahlung
$v_2$ roughly cancel. Thus we find that the total elliptic flow $v^\gamma_2$
of direct photons is roughly the same in cases whether or not jet conversions
are present.
Preliminary data from PHENIX \cite{phenix:07v2} is shown in the left panel.
Both cases are compatible with the data within the large experimental
uncertainties.

To make this study more complete, it is interesting to explore the centrality dependence of the novel double ratio $R^\gamma_{AA}/R^\pi_{AA}$ and elliptic flow of direct photons at RHIC. When the centrality of the collisions increases, the resulting decreased average path length will lead to reduction of both energy loss and jet conversion rates. Hence we expect that the difference  between the scenarios with and without jet conversions for the double ratio $R^\gamma_{AA}/R^\pi_{AA}$ will shrink with increasing centrality. For the elliptic flow of direct photons, the centrality dependence is not so straight forward to estimate because changes in energy loss and jet conversion rates try to sway $v_2$ in different directions. We calculate the elliptic flow of direct photons produced in Au+Au collisions at $\sqrt{s_{NN}}$=200 GeV in the more peripheral centrality bins of 30-40\% and 40-50\%, respectively, and show the results in Fig.\ \ref{v_2_diff_acce}. The resulting total elliptic flow of direct photons is just as small as for the 20-30\% centrality bin. Thus direct photon $v_2$ is not sensitive to the impact parameter in Au+Au collisions due to the cancelation between jet conversions and jet energy loss.

Let us now proceed to discuss the impact of conversions on strange hadrons,
in particular kaons. In Ref.\ \cite{weiliu1} only $u$ and $d$ quarks had been
included alongside gluons, and the resulting $p/\pi^+$ and $\bar p/\pi^-$
ratios reproduce the data from STAR \cite{adams}. We neglect the relatively
small current mass of the strange quark and replace it solely with the thermal
mass $\sim gT$ in the medium when calculating its drag coefficient and
conversion width. This might slightly overestimate conversion rates to
strange quarks, but it simplifies the treatment \cite{weiliu1}.

In the left panel of Fig.\ \ref{raa_k0} we show the spectrum of $K^0_s$ in
$p+p$ collisions calculated as a baseline. On the right hand side
we plot the nuclear modification factor $R^K_{AA}$ of the $K^0_S$ as a
function of transverse momentum $p_T$. We observe that conversions greatly
enhance the yield of $K^0_S$ in nuclear collisions. In fact,
conversions could lead to a $R_{AA}$ that is up to a factor 2 larger at high
$p_T$ than that for pions or protons. Note again, that a recombination
contribution is not included here, so caution has to be exercised at lower
$p_T$. We predict that the measurement of the nuclear modification factor
of kaons and Lambdas at high $p_T$ will provide an unique signal for
jet conversions in the QGP formed at RHIC.

\section{A Look at LHC}
\label{lhc}

At LHC energies jet physics is expected to play an even bigger role than
at RHIC. Using the procedures developed previously we compute the
expected nuclear modification factors $R^\gamma_{AA}$ and $R^K_{AA}$, and the
elliptic flow $v^\gamma_2$ for direct photons as well as various double ratios
in Pb+Pb collisions at $\sqrt{s_{NN}} = 5.5$ TeV. We want to gauge the effect of jet conversions on high-$p_T$ measurements at LHC.


\begin{figure*}[t]
\centerline{
\includegraphics[width=3.0in,height=3.0in,angle=-90]{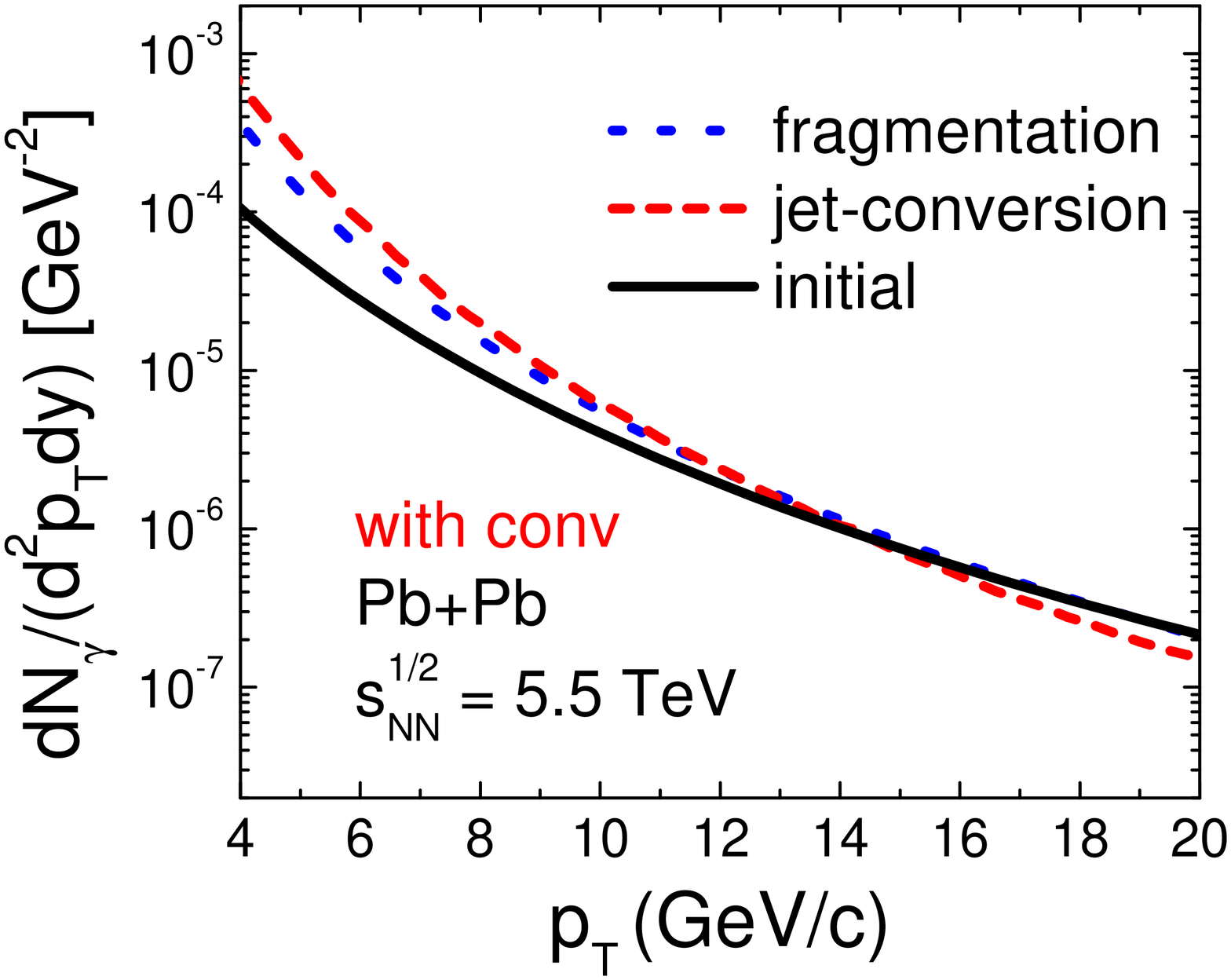}
\hspace{0.2cm}
\includegraphics[width=3.0in,height=3.0in,angle=-90]{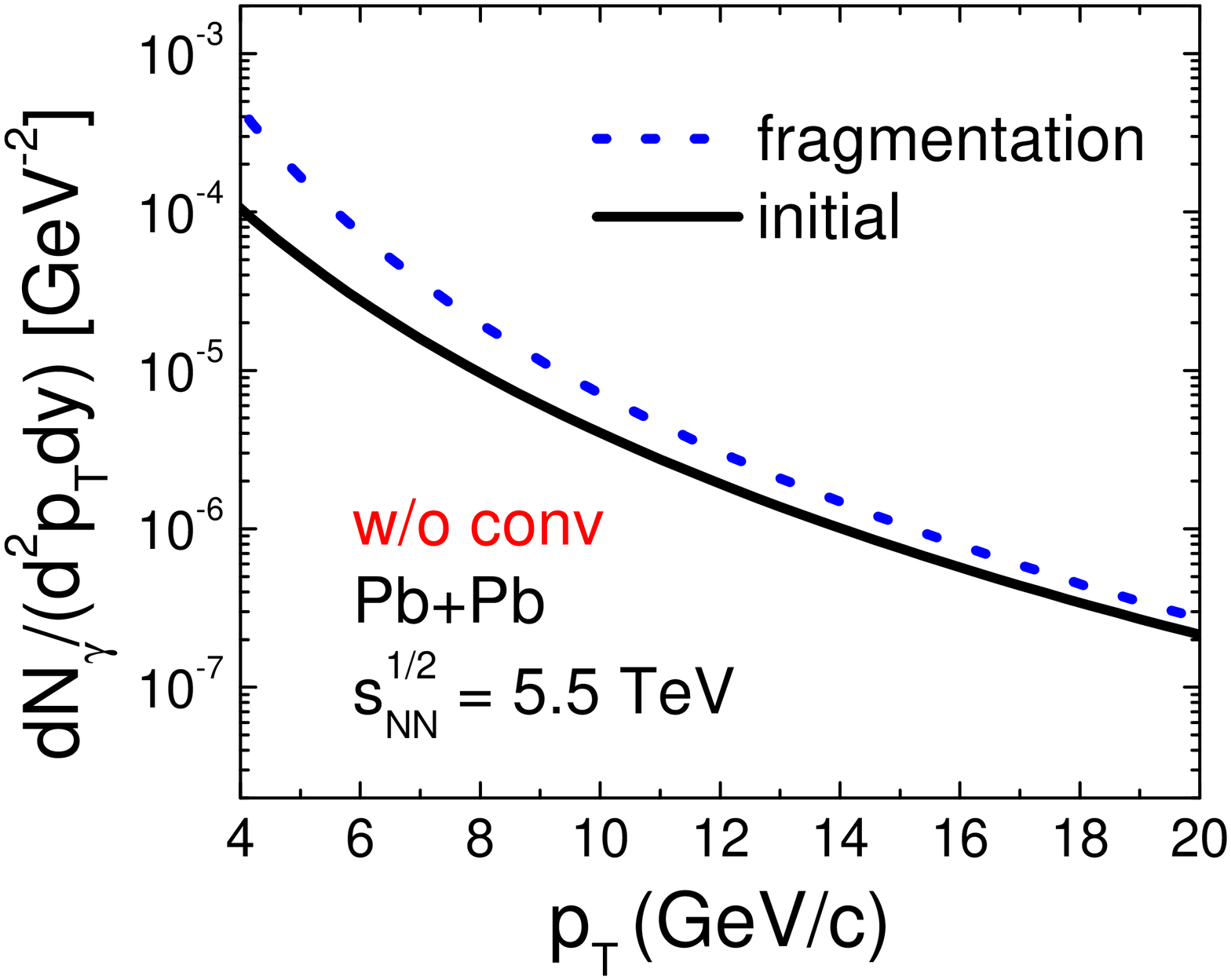}}
\caption{(Color online) Photon spectra in central Pb+Pb collisions at
$\sqrt{s_{NN}}=5.5$ TeV as functions of transverse momentum $p_T$ with
(left panel) or without conversions (right panel).}
\label{lhc_spectra}
\end{figure*}

\begin{figure*}[t]
\centerline{
\includegraphics[width=2.4in,height=2.2in,angle=-90]{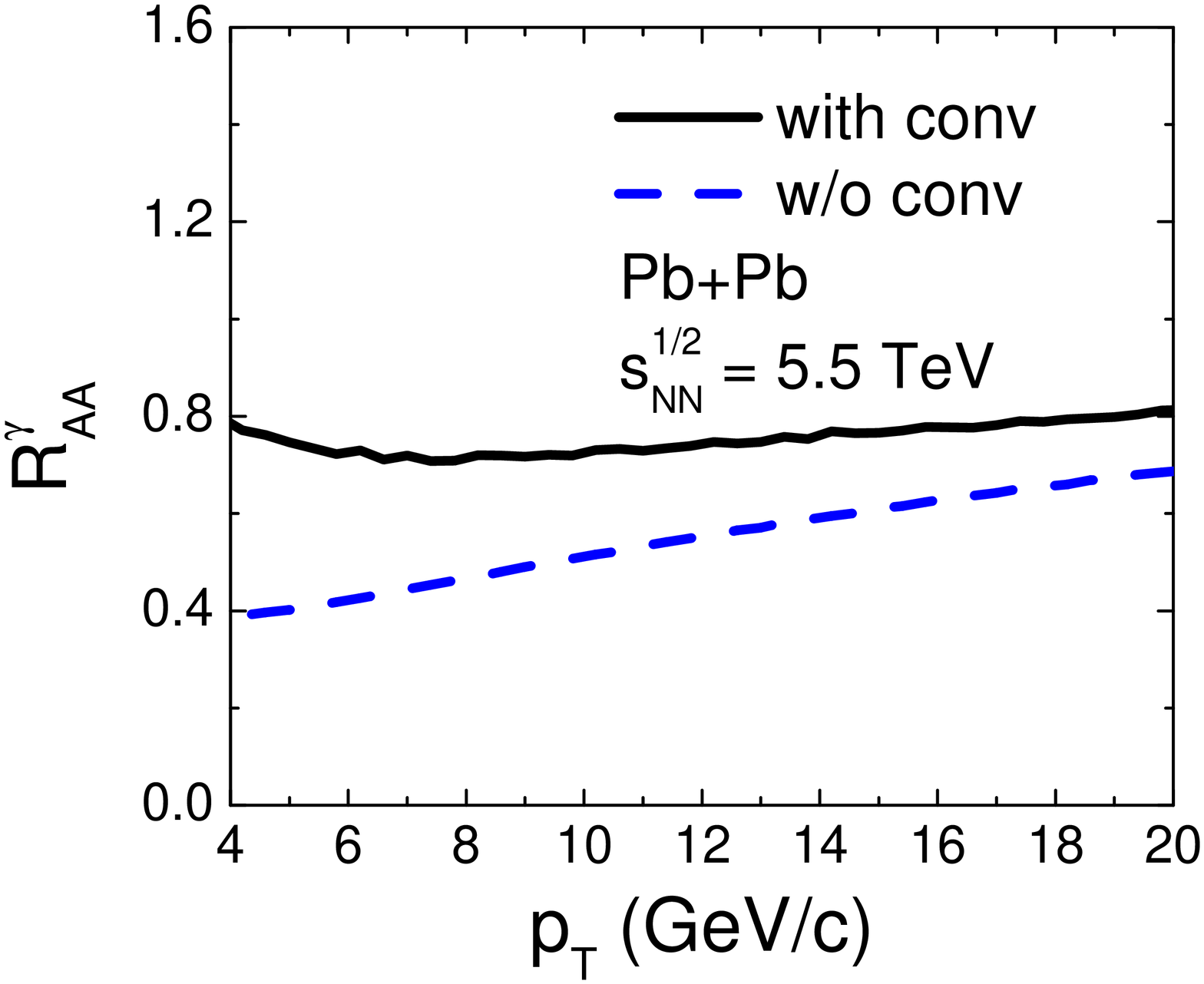}
\hspace{0.2cm}
\includegraphics[width=2.4in,height=2.2in,angle=-90]{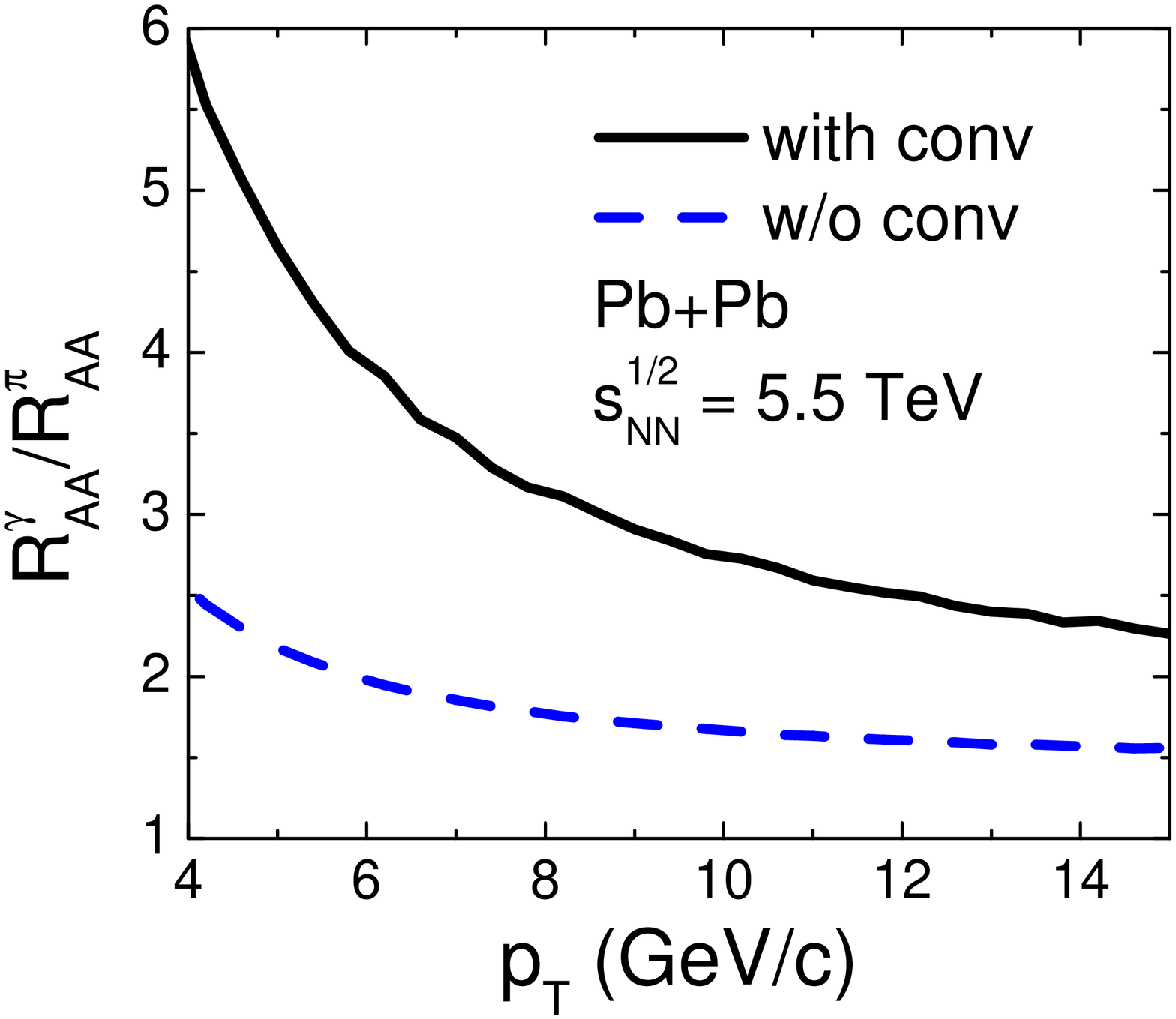}
\hspace{0.2cm}
\includegraphics[width=2.4in,height=2.2in,angle=-90]{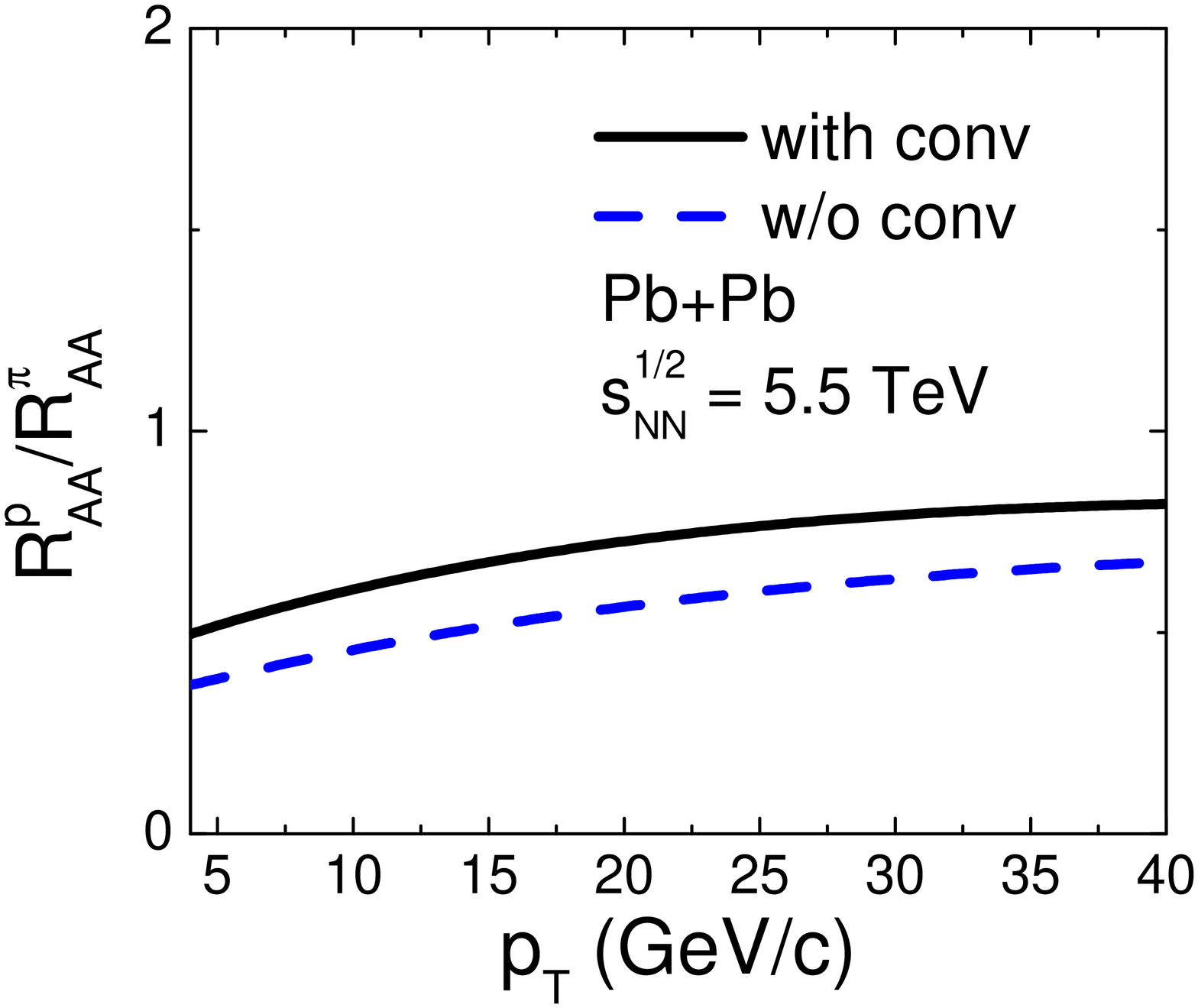}}
\caption{(Color online) Same quantities as those in Fig.\ref{raa} in central Pb+Pb collisions at $\sqrt{s_{NN}}=5.5$ TeV as functions of transverse momentum with or without conversions.}
\label{lhc_raa}
\end{figure*}

We use jet spectra at 5.5 TeV by multiplying those obtained from PYTHIA in $p+p$ collision at the same energy by the number of binary collisions
\cite{weiliu2} ($\langle N_{coll}\rangle\approx$ 1700 \cite{david}).
We also assume that the fireball produced in central collisions at LHC is
cylindrical and Bjorken boost-invariant. The parameters of the fireball are
taken from Ref.\ \cite{zhang} where thermal charm production at LHC was
studied. Specifically, we choose initial proper time $\tau_0=0.2$ fm/$c$ for
the formation of the equilibrated QGP with initial temperature $T_0=700$ MeV,
and flow acceleration $a=0.1$ $c^2$/fm. This yields reasonable final
transverse flow velocities with initial radius $R_0=7.0$ fm. Fixing the
parameters at the critical temperature $T_c=175$ MeV via entropy conservation we obtain $\tau_c=6.3$ fm/c. We apply the same test particle
Monte Carlo method as above with the same set of $K$ factors.

Our results for direct photons are shown in Fig.\ \ref{lhc_spectra} for $K=4$
(1 for photons) (left panel) and $K=0$ (right panel), respectively.
Direct photons
from jet-photon conversions and jet fragmentation dominate below $p_T = 16$
GeV/$c$ when conversions are switched on, with photons from conversion being
the largest source below 10 GeV/$c$. Without jet conversions the direct
photons from jet fragmentation are always brighter than those from initial
hard collisions up to 20 GeV/$c$, because of the reduced conversion of quarks.

It is straightforward to calculate the nuclear modification factor
$R^\gamma_{AA}$, which is shown in the left panel of Fig.\ \ref{lhc_raa}. The
difference between both scenarios can be as large as a factor of 2 at $p_T =
4$ GeV/$c$, and decreases to a factor of about 1.25 at 20 GeV/$c$. We also
observe a large increase in the double ratio $r_{\gamma/\pi^+} = R^\gamma_{AA} /
R^{\pi^+}_{AA}$ if conversions are included, just as seen for RHIC as well. For comparison, we also show the result of the double ratio $r_{p/\pi^+}$ at LHC .


\begin{figure*}[ht]
\centerline{
\includegraphics[width=3.0in,height=3.0in,angle=-90]{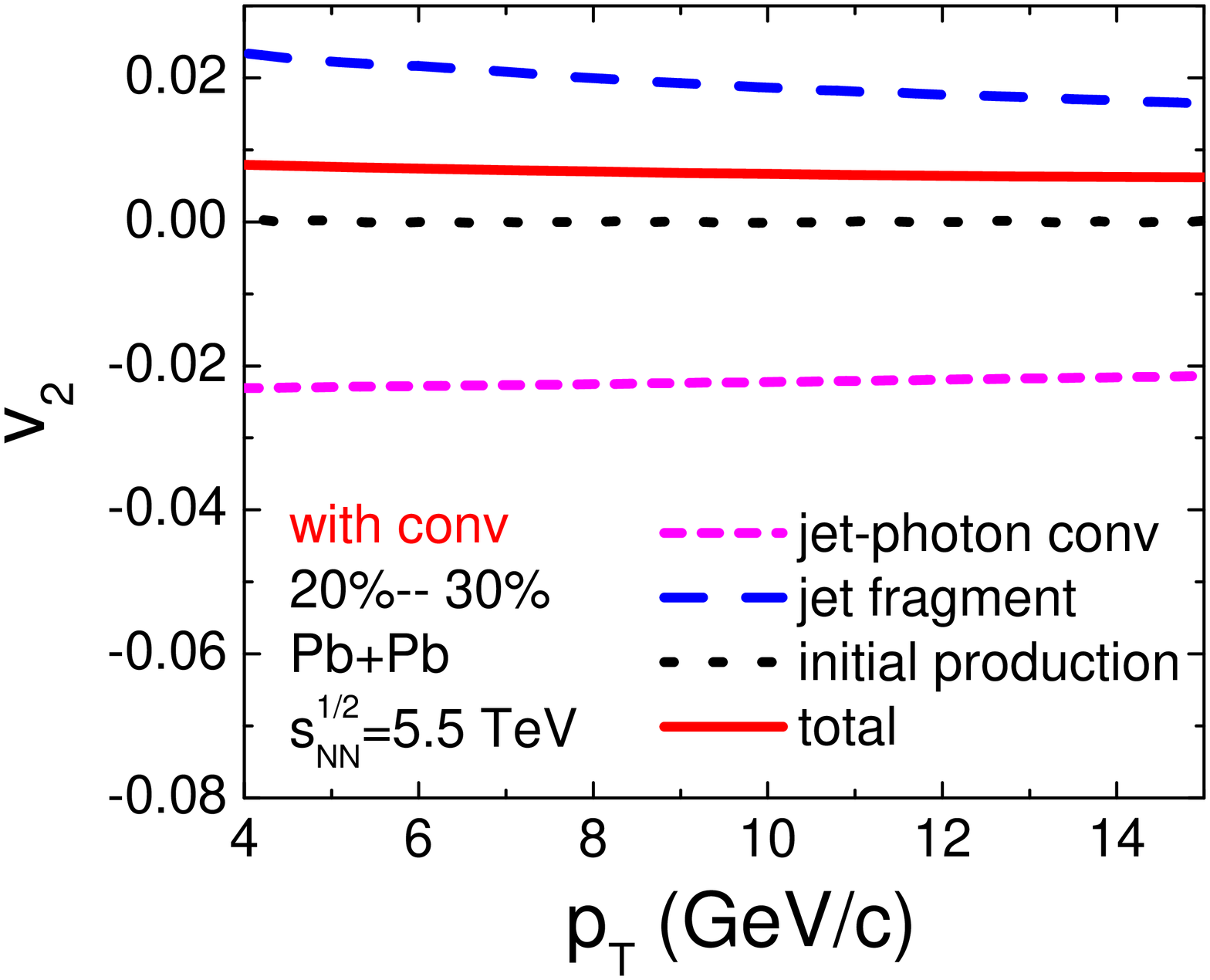}
\hspace{-0.2cm}
\includegraphics[width=3.0in,height=3.0in,angle=-90]{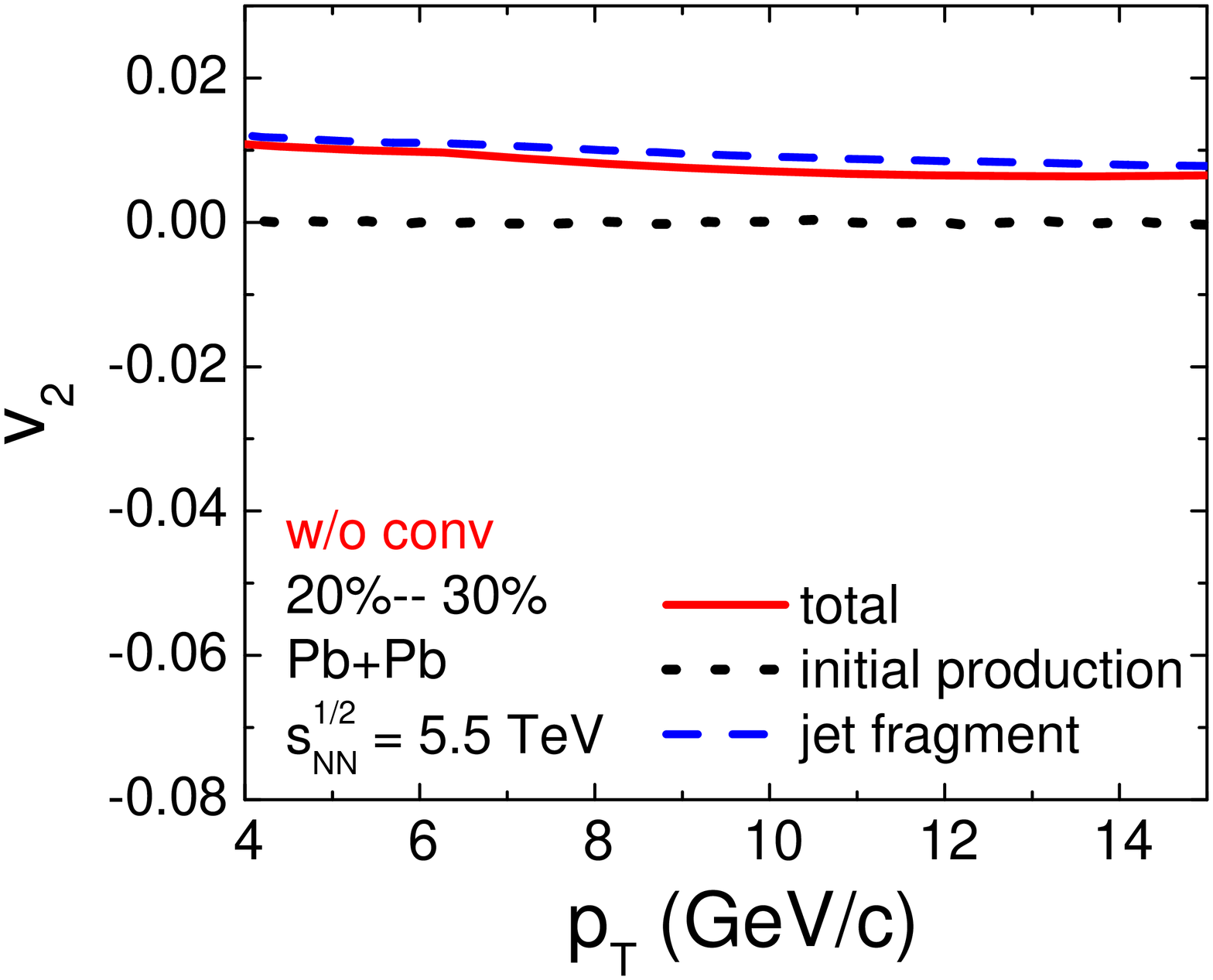}}
\caption{(Color online) Elliptic flow $v^\gamma_2$ of direct photons in
  peripheral Pb+Pb collisions with centrality of $20 -30\% $ at $\sqrt{s_{NN}}=5.5$ TeV as functions of transverse momentum with (left panel) or without conversions (right panel).}
\label{lhc_v_2}
\end{figure*}

\begin{figure*}[t]
\includegraphics[width=3.0in,height=3.5in,angle=-90]{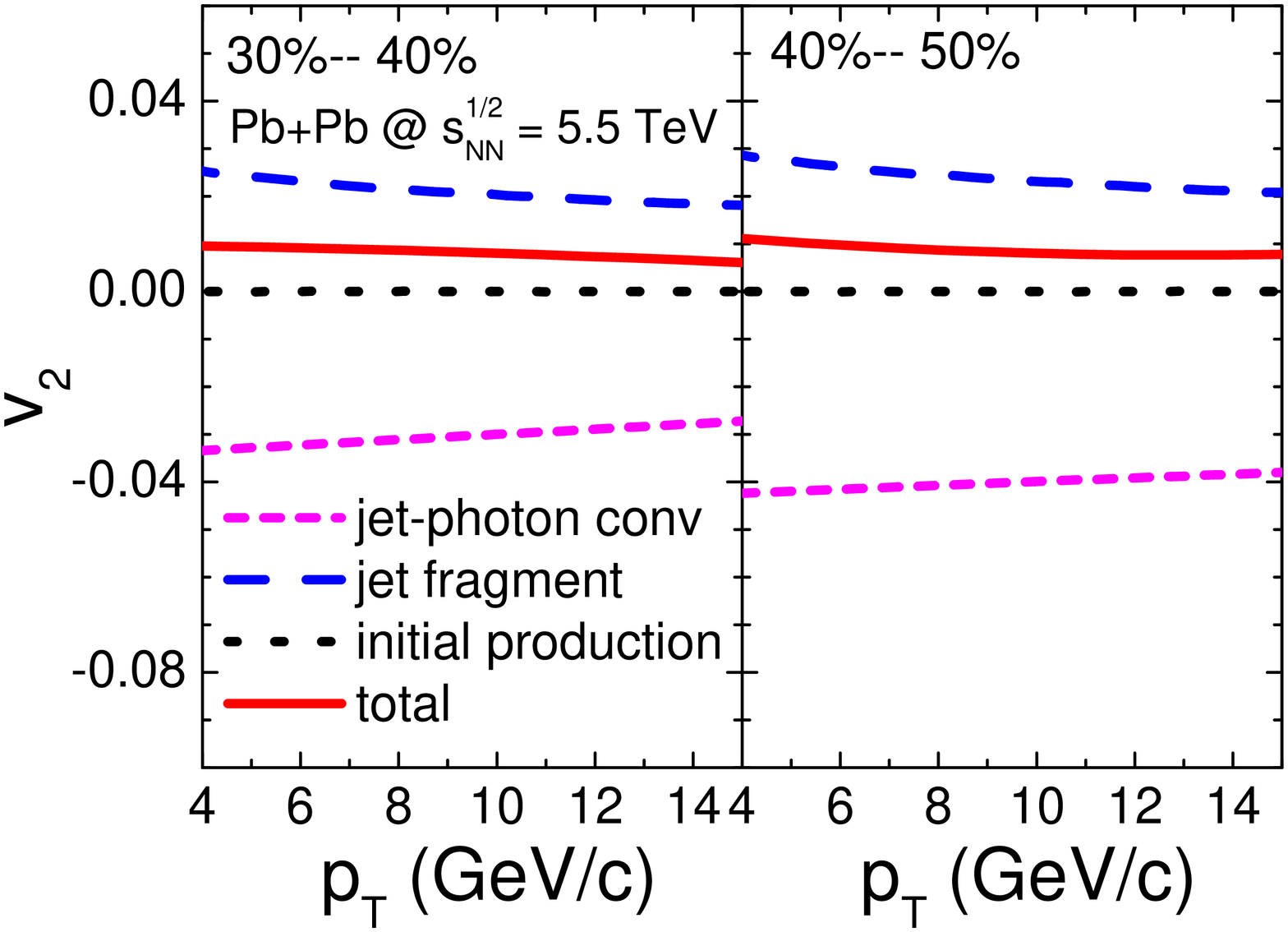}
\caption{(Color online) Elliptic flow $v^\gamma_2$ of direct photons in
  peripheral Pb+Pb collisions with centrality of $30 -40\% $ (left) and $40 - 50\% $ (right) at $\sqrt{s_{NN}}=5.5$ TeV as functions of transverse momentum.}
\label{lhc_v_2_diff_acce}
\end{figure*}

Since we have seen that direct photons at LHC are mostly from
jet-photon conversions and final state jet fragmentation, they are
strongly influenced by jet conversions. Photon elliptic flow
$v^\gamma_2$ might reveal those large effects. The parameters of the fireball formed in peripheral collisions at LHC are extracted from information about centrality, number of participants, and number of binary collisions in Pb+Pb collisions at $\sqrt{s_{NN}}$= 5.5= TeV \cite{david}. For simplicity, in collisions at LHC within the centrality bin $20-30$\%, we use the same geometric parameters (the length of the long and short axes) for the initial fireball as at RHIC, but we set the initial temperature and proper time of QGP formation to $T_0$ = 700 MeV and $\tau_0$ = 0.2 fm/c, respectively, consistent with central collisions. The time dependence of the temperature is also fixed via entropy conservation.

In Fig.\ \ref{lhc_v_2} we show the resulting elliptic flow $v^\gamma_2$ of
direct photons in Pb+Pb collisions at $\sqrt{s_{AA}}$ = 5.5 TeV with and
without conversions. The elliptic flow from photon
fragmentation off jets is positive with a magnitude of ~2\% in the
case of jet conversions, but it is reduced by a factor of 2 in the case of
missing jet conversions due to the reduced quenching for quark.
The resulting total elliptic flow of direct photons without jet
conversions differs only slightly from that of jet fragmentation alone, since
direct prompt photons from the initial state are subdominant below 20
GeV/$c$. With jet conversions included, we again observe a cancellation
between the negative anisotropy from jet-photon conversions and the increase
in jet fragmentation with positive anisotropy, leading to a total elliptic
flow $v^\gamma_2$ of almost the same magnitude as that without jet
conversions. Together with our previous result at RHIC energies this might
indicate that photon elliptic flow, despite its early promise, might need
very large experimental precision to be used as a probe.
The situation might be much improved, however, if measurements with additional
isolation cuts could be made \cite{simon1}.

We also predict the centrality dependence of the elliptic flow for direct photons in Pb+Pb collisions at $\sqrt{s_{NN}}$=5.5 TeV and show the results in Fig.\ \ref{lhc_v_2_diff_acce}. The numerical results confirm once more that the total elliptic flow of direct photons at high transverse momentum varies only weakly with the centrality of the collision.

\begin{figure}[ht]
\centerline{
\includegraphics[width=3.0in,height=3.0in,angle=0]{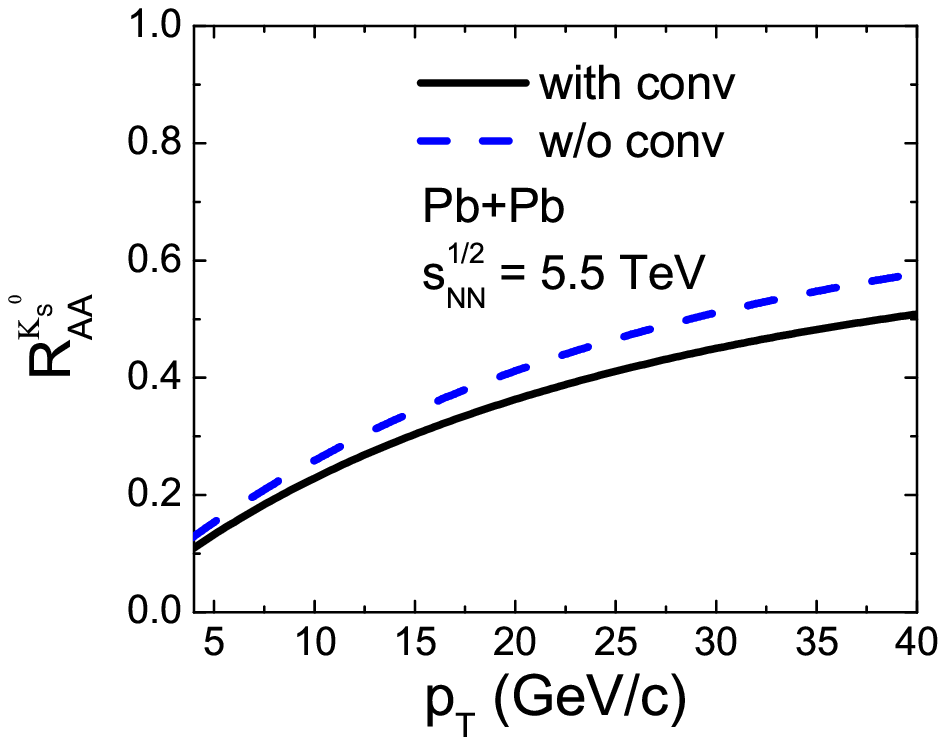}}
\caption{(Color online) Nuclear modification factor $R_{AA}$ for $K^0_S$ in $Pb+Pb$ collisions at $\sqrt{s_{NN}}$=5.5 TeV as function of transverse momentum.}
\label{lhc_raa_k0}
\end{figure}

In Fig.\ \ref{lhc_raa_k0} we show $R^K_{AA}$ of $K^0_S$ in central collisions
at LHC. We observe that jet conversions have only a small impact on
$R^K_{AA}$. This had to be expected, because initial jet production is
gluon dominated and strange quark jets are not very much suppressed to
begin with, very unlike the situation found at RHIC.
The impact of jet conversions on the $R_{AA}$ of kaons is similar to the
effect on pions at LHC \cite{weiliu3}. We conclude that strange hadrons
might not be a good probe of jet conversions any more at LHC energies.

\section{Summary and Discussion}\label{summary}

We studied the impact of flavor conversions of jets on observables in
heavy ion collisions. We argued that the relaxation of jet flavors
toward chemical equilibrium can be an independent measure for the
strength of the jet-medium coupling. We presented a computation of
several observables which are sensitive to conversion
processes at RHIC and LHC using a Fokker-Planck approach and coupled rate
equations for conversions. Rates were based on leading order cross sections
with a $K$ factor.

We find a large change in the nuclear modification factor $R^\gamma_{AA}$
when jet conversions are switched on, while the effects on photonic elliptic
flow $v^\gamma_2$ tend to cancel.  We find our results to be consistent with
the available experimental data from PHENIX \cite{adler2}. We also calculated
the impact of jet conversions on the double ratios of nuclear modification
factor $r_{\gamma/\pi^+} = R^\gamma_{AA}/R^{\pi^+}_{AA}$ and $r_{p/\pi^+} = R^p_{AA}/R^{\pi^+}_{AA}$ and find that these quantities are quite sensitive to the
presence of jet conversions. The difference can be as large as a factor of two.
At the same time such double ratios might be extracted from experimental data
with rather small systematic error bars. Large $K$-factors in
our leading order treatment would hint at non-perturbative mechanisms, or
to the presence of additional effects, like the influence of increased
multiplicities in modified jets \cite{Sapeta:2007ad}.
We also advocate the measurement of strange hadrons at high $p_T$ as a
possible signature for the jet-medium coupling. The low initial strange
quark content of jets at RHIC makes kaons and other strange hadrons a
a very sensitive probe for the coupling to the medium.

\begin{acknowledgments}
We thank Che-Ming Ko for useful discussions and suggestions.
This work was supported by RIKEN/BNL, DOE grant DE-AC02-98CH10886, and the
Texas A\&M College of Science.
\end{acknowledgments}

\end{document}